\documentclass[useAMS,usenatbib]{mn2e}

\usepackage{graphicx}
\usepackage{times}
\usepackage{natbib}
\usepackage{color}
\usepackage{todonotes}

\newcommand{\apjs}{ApJS}

\newcommand{\apjl}{ApJL}

\newcommand{\aap}{A \& A}

\newcommand{\enzo}{\it{\small ENZO}}

\begin{document}
 
\title[Multi wavelength cross-correlation]{Multi wavelength cross-correlation analysis of the simulated cosmic web}
\author[C. Gheller, F. Vazza]{C. Gheller$^1$\thanks{E-mail: claudio.gheller@gmail.com}, F. Vazza$^{4,3,2}$\thanks{E-mail: franco.vazza2@unibo.it}\\
$^{1}$ Swiss Plasma Center, EPFL, SB SPC Station 13 - 1015 Lausanne, Switzerland\\
$^{2}$Istituto di Radio Astronomia, INAF, Via Gobetti 101, 40121 Bologna, Italy\\
$^{3}$ Hamburger Sternwarte, Gojenbergsweg 112, 21029 Hamburg, Germany\\
$^{4}$ Dipartimento di Fisica e Astronomia, Universit\'{a} di Bologna, Via Gobetti 92/3, 40121, Bologna, Italy
}

\maketitle
\begin{abstract}

We used magneto-hydrodynamical cosmological simulations to investigate the  cross-correlation between different observables (i.e. X-ray emission, Sunyaev-Zeldovich signal at 21 cm,  HI temperature decrement, diffuse synchrotron emission and Faraday Rotation) as a probe of the diffuse matter distribution in the cosmic web. We adopt an uniform and simplistic approach to produce synthetic observations at various wavelengths, and we compare the detection chances of different combinations of observables correlated with each other and with the underlying galaxy distribution in the volume. With presently available surveys of galaxies and existing instruments, the best chances to detect the diffuse gas in the cosmic web outside of halos is by cross-correlating the distribution of galaxies with Sunyaev-Zeldovich observations. We also find that the cross-correlation  between the galaxy network and the radio emission or the Faraday Rotation can already be used to limit the  amplitude of extragalactic magnetic fields,  well outside of the cluster volume usually explored by existing radio observations, and to probe the origin of cosmic magnetism with the future generation of radio surveys.

\end{abstract}

\label{firstpage} 

\section{Introduction}
\label{sec:intro}

Galaxy surveys and numerical simulations have consistently shown that the large scale structure of the Universe is organised in a hierarchy of highly overdense halos, mildly overdense filaments and underdense voids.  A large fraction of the baryonic matter (around 50\%) should indeed reside in the form of plasma in such cosmic web, at densities 10-100 times the average cosmic value and temperature mostly of $10^5-10^7$K, forming the Warm-Hot Intergalactic Medium (WHIM, \citealt{1999ApJ...514....1C}). 

The detection and the characterisation of the WHIM in cosmic filaments is of primary interest. Firstly, finding in the WHIM the missing baryonic mass \citep[e.g.][]{2016xnnd.confE..27N} would verify one of the pillars of modern cosmological structure formation paradigm. Its distribution would trace the geometry and define the topology of the universe \citep[e.g.][]{2014MNRAS.441.2923C}. Furthermore, filaments evolve with adiabatic physics (besides gravity) driving the gas dynamics, but with other physical processes possibly influencing their chemical \citep[e.g.][]{2019MNRAS.486.3766M} and magnetic \citep[e.g.][]{gv19} properties. Thanks to a less violent growth compared to galaxy clusters, they preserve essential information on the original environment in which structure formation takes place, besides on magnetogenesis and primordial magnetism.


Observing the WHIM has been so far a challenge at all wavelengths \citep[e.g.][]{2012Natur.487..202D,2015A&A...583A.142N,2018ApJ...867...25C,2019arXiv190910518C}, due  to its extremely low particle density, leading to a faint emission at the limit or below the current instrumental sensitivity, which is also strongly affected by background/foreground contributions and observational noise and artefacts. Only since recently imaging of the WHIM, connected to massive nearby galaxy clusters \citep[][]{2015Natur.528..105E} or to dense proto-clusters at high redshift \citep[][]{2019arXiv191001324U}, has become feasible. A few additional potential detections of filaments around cosmic structures have been reported, using the Sunyaev Zeldovich effect \citep[e.g.][]{2013A&A...550A.134P,Tanimura19} or radio waves \citep[e.g.][]{2018MNRAS.tmp.1093V,2018MNRAS.478..885B,2019Sci...364..981G}.
More recently, \citet{2019A&A...625A..67T} and \citet{2019A&A...624A..48D} have  presented the possible first detection of the coldest part of the coldest part of the WHIM in the cosmic web,   by stacking the thermal Sunyaev Zeldovich signal of pairs of galaxies or pairs of galaxy groups.

Statistical techniques can be adopted to detect the presence of the cosmic web, overcoming the limits of current instruments.
In this work we will rely on the {\it cross-correlation} analysis to measure the signatures of large-scales diffuse emission considering various  ``mass tracers'' (galaxies, X-ray emission, radio emission etc.). Cross-correlation is a widely used methodology in signal and image processing, that we will exploit in order to identify faint signals, below the sensitivity of single instruments. 
In fact noise and, generally speaking, artefacts and background/foreground affecting different types of observation are completely unrelated from each other and, in general, from the signal, hence their combination tends to cancel out. On the other hand, actual signals coming from the same source tend to sum constructively and magnify. 


Many examples of successful applications of the cross-correlation analysis in astronomy can be found, the following list far from being exhaustive. The method has been first used to detect the Integrated Sachs-Wolfe effect in Cosmic Microwave Background (CMB) data \citep[e.g.][]{2004ApJ...608...10N}. 
\citet{2019A&A...625L...4H} detected the cross-correlation between X-rays and CMB weak lensing, as well as performed auto- and cross-correlation of SZ, X-rays, and weak-lensing  to assess the galaxy cluster hydrostatic mass bias. 
\citet{2016MNRAS.456.1495S} investigated the detectability of the cross-correlatation between galaxy distribution, SZ and X-ray for various future surveys and with the goal of detecting the circumgalactic medium, estimating a maximum detection efficiency for $\sim 10^{13} M_{\odot}$ halos at $z \sim 1-2$. 
\citet{2018PhRvD..98j3518M} proposed the detection of the WHIM by cross correlating the Dispersion Measure of Fast Radio Bursts and thermal SZ maps, which is now a concrete possibility thanks to the deployment of dedicated instruments for Fast Radio Burst (e.g. CHIME). 
Moving to higher redshift, \citet{2018MNRAS.480...26M} computed the amplitude of the  cross-correlation signal between the emission from energetic and high-z source of X-ray background and the HI signal from intergalactic gas the epoch of reionization.

In the radio domain, cross-correlation has been adopted to detect faint, spatially correlated, emission below the noise limit of radio surveys.
\citet{vern17} and \citet{brown17} have presented first attempts of cross-correlating the distribution of radio emission in the continuum to that of galaxies in large portions of the sky, seeking for a positive correlation between diffuse emission and the cosmic web.
\citet{vern17} cross-correlated early MWA observations at $169$ MHz with the distribution of galaxies in the WISE and 2MASS galaxy surveys, for a $22^\circ \times 22^\circ$ field of view. 
\citet{brown17} used instead $2.3 \rm ~GHz$ observations from the S-PASS survey, and cross-correlated them with template radio emission from a constrained MHD simulation. 
In both cases, no significant detection of a cross-correlation was found, probably due to the limited sensitivity and spatial resolution of radio data (as well as due to the possible contamination of unresolved radiogalaxies, whose clustering properties correlate with that of the galaxy distribution).
However, upper limits on the average amplitude of magnetic fields in the cosmic web in the range $B \leq 0.01-0.1 ~\rm \mu G$ were derived from both works. \\


Being a statistical methodology, the cross-correlation has the limitation that it provides no direct inference of the underlying matter distribution and of the physical mechanisms behind the detected signal. Proper modelling is necessary to link the observed statistics to the properties of the corresponding sources.

Numerical simulations represent the most effective and general tool to pursue such objective. We have exploited a number of Magneto-Hydrodynamical (MHD) simulations run by our group, encompassing a broad variety of magnetic, astrophysical and galaxy evolution setup to create a variety of mock observations. From the simulations, in fact, different types of signals (synchrotron, SZ, X...) can be calculated and correlated with each other or with the dark matter, representative of the galaxy distribution. The mock observations have been generated with or without the contribution of additional noise. The latter allows to better discriminate between different prescriptions for the gas physics and its magnetic properties. Random noise has been added to images considering the typical detection threshold of several instruments in the relevant energy bands. We have considered both instruments ``currently'' available, meaning that they are already operational (like ASKAP or LOFAR-HBA) or they will be in the next future (like eROSITA), and ``future'' instruments, that will become available on a longer time horizon (like SKA or ATHENA-WFI). The former points out the expected detection with data immediately available, the latter the possible improvements in the long-term perspective.

The resulting simulated dataset has then be used to perform a first systematic survey of the degree of cross-correlation measured between several relevant observable signatures of the diffuse gas in the cosmic web. Such survey is intended both to guide future studies adopting the cross-correlation statistics to detect the WHIM in various kind of observations, and to interpret possible detections in terms of the properties of the underlying  diffuse gas component. We present a first example of such kind of interpretation, by comparing the cross-correlation measured between galaxies and synchrotron emission in our different models with the results obtained by \citet{vern17}.\\



Our paper is organised as follows. The numerical methods used for the cosmological simulations and to generate the multi-wavelengths mock observations will be described in Section 2. In the same Section we will also validate the different models assessing their reliability for the performed analysis. In Section 3 we will give an essential introduction to cross-correlation and its usage on the simulated images. In Section 4 the results of applying the cross-correlation analysis to the different models will be presented and discussed. Conclusions will be drawn in Section 5.

\section{Simulations and mock observations}
\label{sec:data}


\subsection{Numerical simulations}
\label{sec:simulations}

Our simulations adopted the cosmological Eulerian code  {\enzo} \citep{enzo14}, with  a fixed mesh resolution. The code has been customised by our group mainly with the purpose of including different mechanisms for the seeding of magnetic fields in cosmology, as explained in detail in \citet{va17cqg} and \citet{gv19}. 

The magneto-hydrodynamics (MHD) solver used in our simulations implements the conservative Dedner formulation \citep[][]{ded02}, which utilises hyperbolic divergence cleaning to keep the $\nabla \cdot \vec{B}$ as small as possible, and the 
Piecewise Linear Method reconstruction (PLM) technique with fluxes calculated using the Harten-Lax-Van Leer (HLL) approximate Riemann solver. Time integration is performed using the total variation diminishing (TVD) second-order Runge-Kutta (RK) scheme \citep[][]{1988JCoPh..77..439S}.  We used the GPU-accelerated MHD version of {\enzo} by \citet[][]{wang10}, which gives a $\sim \times 4$ speedup compared to the more standard CPU version of {\enzo} in the $1024^3$ uniform grid runs used here. Constant spatial resolution has the advantage of providing the best resolved description of magnetic fields even in low-density regions, which would typically go unrefined by Adaptive Mesh Refinement (AMR) approaches. 

We have exploited a subset of data extracted from the ``Chronos++ suite'' {\footnote{http://cosmosimfrazza.myfreesites.net/the\_magnetic\_cosmic\_web}}, which in total includes $24$ different models, designed to explore various plausible scenarios for the origin and evolution of extra galactic magnetic fields  (\citep[][]{va17cqg}). 
Here we focus on three models: {\it primordial}, {\it dynamo} and {\it astrophysical} scenarios (see also Tab. 1{\footnote {For consistency with the nomenclature used in our previous works on this topic, here we adopted the same ID of models used elsewhere.}}): 

\begin{itemize}
\item{\it primordial} model: a non-radiative simulation in which we assumed the existence of a volume-filling magnetic fields at the beginning of the simulation, with magnitude $B_0 = 1$ nG. This simulation represents our ``baseline" reference model for the cosmic web.  
\item {\it dynamo} model: also a non-radiative simulation as before, in which we estimated at run-time via sub-grid modelling the small-scale dynamo amplification of very weak seed field of primordial origin  ($B_0 = 10^{-9} \rm nG$). 
Here the dissipation of solenoidal turbulence into magnetic field amplification is estimated on the fly  by extrapolating the information resolved at our fixed $83.3 ~\rm kpc$ cell resolution. In this work we consider model "DYN5" of \citet{gv19}, which gives a reasonable match to the magnetic field strength in galaxy clusters.  
\item {\it astrophysical} model: a more sophisticated simulation including radiative gas cooling, chemistry, star formation and thermal/magnetic feedback from  a) stellar activity and/or b) feedback by supermassive black holes (SMBH), simulated at run time using prescriptions available in {\enzo}  \citep[e.g.][]{2003ApJ...590L...1K,2011ApJ...738...54K,enzo14}.  Our reference model here, ``CSFBH2", assumes accretion for SMBH following from the  spherical Bondi-Hoyle formula with a fixed $0.01 ~\rm M_{\odot}/yr$ accretion rate, and a fixed "boost" factor to the mass growth rate of SMBH ($\alpha_{\rm Bondi} =1000$) to balance the effect of coarse resolution, properly resolving the mass accretion rate onto our simulated SMBH particles. We have extended {\enzo} coupling thermal feedback to the injection of additional magnetic energy via bipolar jets, with an efficiency with respect to the feedback energy computed at run-time, $\epsilon_{\rm SF,b}$ and $\epsilon_{\rm BH,b}$ for the stellar and SMBH, respectively. Here we used $\epsilon_{\rm SF,b}=10\%$ and $\epsilon_{\rm BH,b}=1\%$ for the magnetic feedback, while for the feedback efficiency (referred to the $\epsilon_{\rm SF, BH} = \dot{M} c^2$ energy accreted by star forming or black hole particles) we used $10^{-8}$ and $0.01$ respectively. This run gives a good match to the observed cosmic average star formation rate as well as to observed galaxy cluster scaling relations (as discussed in Section \ref{sec:validation} and in \citealt{va17cqg}). The implementation of cooling adopted here follows the non-equilibrium evolution of primordial (metal-free) gas. The chemical rate equations are solved using a semi-implicit backward difference scheme, while heating and cooling processes include a number of processes (e.g. atomic line excitation, recombination, collisional excitation, free-free transitions, Compton scattering of the cosmic microwave background and photoionization from metagalactic UV backgrounds). The species that are tracked at run-time in the simulation are only atomic species (i.e. H, H+, He, He+, He++, and electrons), and their evolution is computed by solving the rate equations with one Jacobi iteration with implicit Eulerian time discretization, with a coupling between thermal and chemical states at subcylces in the hydrodynamical timestep \citep[see][and references therein for more details]{enzo14}.  

\end{itemize}


\begin{table*}
\begin{center}
\caption{Main parameters of the three runs in the Chronos++ suite used in the present work. From left to right, columns defines: the presence of radiative cooling or star forming particles, the critical gas number density $n_*$ to trigger star formation in the \citet{2003ApJ...590L...1K} model, the time-scale for star formation $t_*$, the thermal feedback efficiency and the magnetic feedback efficiency ($\epsilon_{\rm SF}$ and $\epsilon_{\rm SF,b}$) from star forming regions; the efficiency of Bondi accretion $\alpha_{\rm Bondi}$ in the \citet{2011ApJ...738...54K} model for SMBH; the thermal feedback efficiency and the magnetic feedback efficiency ($\epsilon_{\rm BH}$ and $\epsilon_{\rm BH,b}$) from SMBH; the intensity of the initial magnetic field, $B_0$; the presence of sub-grid dynamo amplification at run time; the ID of the run and some additional descriptive notes.  All simulations evolved a $85^3 \rm Mpc^3$ volume using $1024^3$ cells and dark matter particles, starting at redshift $z=38$. The name convention of all runs is consistent with \citet{va17cqg}.}
\footnotesize
\centering \tabcolsep 2pt
\begin{tabular}{c|c|c|c|c|c|c|c|c|c|c|c|c|c}
  cooling & star form. & $n_*$ & $t_*$ & $\epsilon_{\rm SF}$ & $\epsilon_{\rm SF,b}$  & $\alpha_{\rm Bondi}$ & $\epsilon_{\rm BH}$ &$\epsilon_{\rm BH,b}$  & $B_0$   & dynamo & ID & description \\ 
  & & [$1/\rm cm^3$]  & [$\rm Gyr$]  &  & & &  &  &  [G] & &  & & \\  \hline \hline
   n &n& - & -& -& -& -  &-& -& $10^{-9}$ & no &  Baseline & primordial,uniform, \\   
   n &n& -  & -& -& -& - &-&  -&$10^{-18}$  & $10 \cdot \epsilon_{\rm dyn}(\mathcal{M})$  & DYN5 & low primordial, efficient dynamo\\
   y &y&0.0001 &1.5 & $10^{-8}$ & 0.01 & $10^3$ fix. &0.05 & 0.01 &   $10^{-18}$  & -  & CSFBH2 & star formation, BH, constant $(\frac{0.01 M_{\odot}} {\rm yr} )$ \\
  \end{tabular}
  \end{center}
\label{table:tab1}
\end{table*}

All the runs adopted the $\Lambda$CDM cosmology, with density parameters $\Omega_{\rm BM} = 0.0478$ (BM representing the Baryonic Matter), $\Omega_{\rm DM} = 0.2602$ (DM being the Dark Matter),  $\Omega_{\Lambda} = 0.692$ ($\Lambda$, being the cosmological constant) and a Hubble constant $H_0 = 67.8$ km/sec/Mpc and $\sigma_8=0.815$ \citep[][]{2016A&A...594A..13P}.  The initial redshift is $z=38$, the spatial  resolution is $83.3 ~\rm kpc/cell$ (comoving) and the constant mass resolution of  $m_{\rm DM}=6.19 \cdot 10^{7}M_{\odot}$ for dark matter particles. Additional details on our sample of simulations can be found in  \citet{va17cqg} and \citet{gv19}.

\begin{figure}
\includegraphics[width=0.49\textwidth]{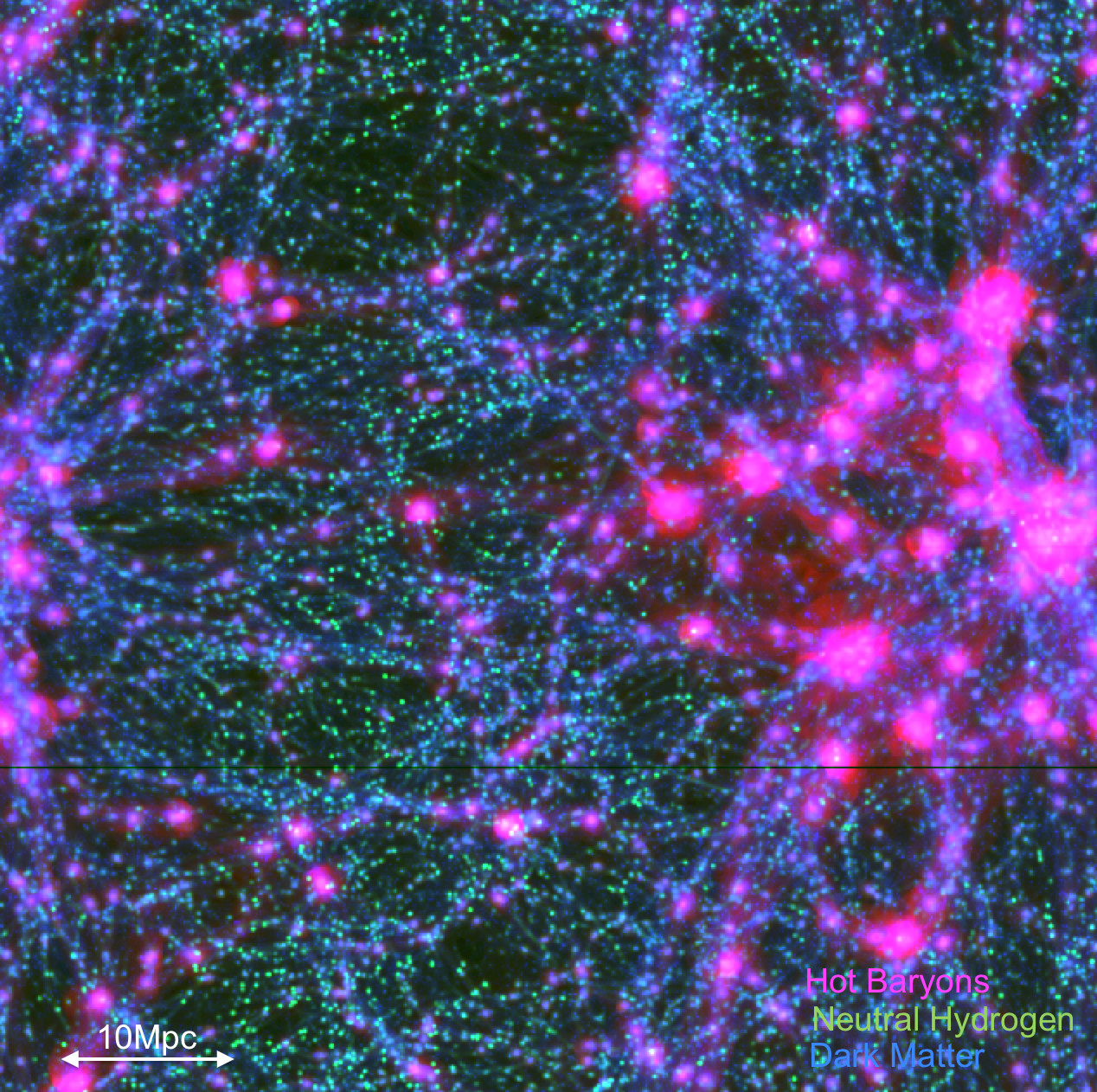}
\caption{RGB rendering of the matter distribution in our simulated volume for the CSFBH2 model at $z=0.045$: dark matter density (blue), ionised gas (pink) and neutral Hydrogen (green).}
\label{fig:rgb}
\end{figure}

An RGB rendering of the projected distribution of different mass components (dark matter, ionised gas and neutral Hydrogen) in our CSFBH2 model at $z=0.045$ is shown in Figure \ref{fig:rgb}. This example shows, at least in a qualitative way, the general difficulty of detecting the diffuse gas in the cosmic via cross-correlation: although on large scales all tracers are well correlated and part of the same web pattern, on scales smaller than a few $\rm Mpc$ they present different level of clustering and concentration, making their cross-correlation signal challenging to detect.

\begin{figure}
\includegraphics[width=0.45\textwidth]{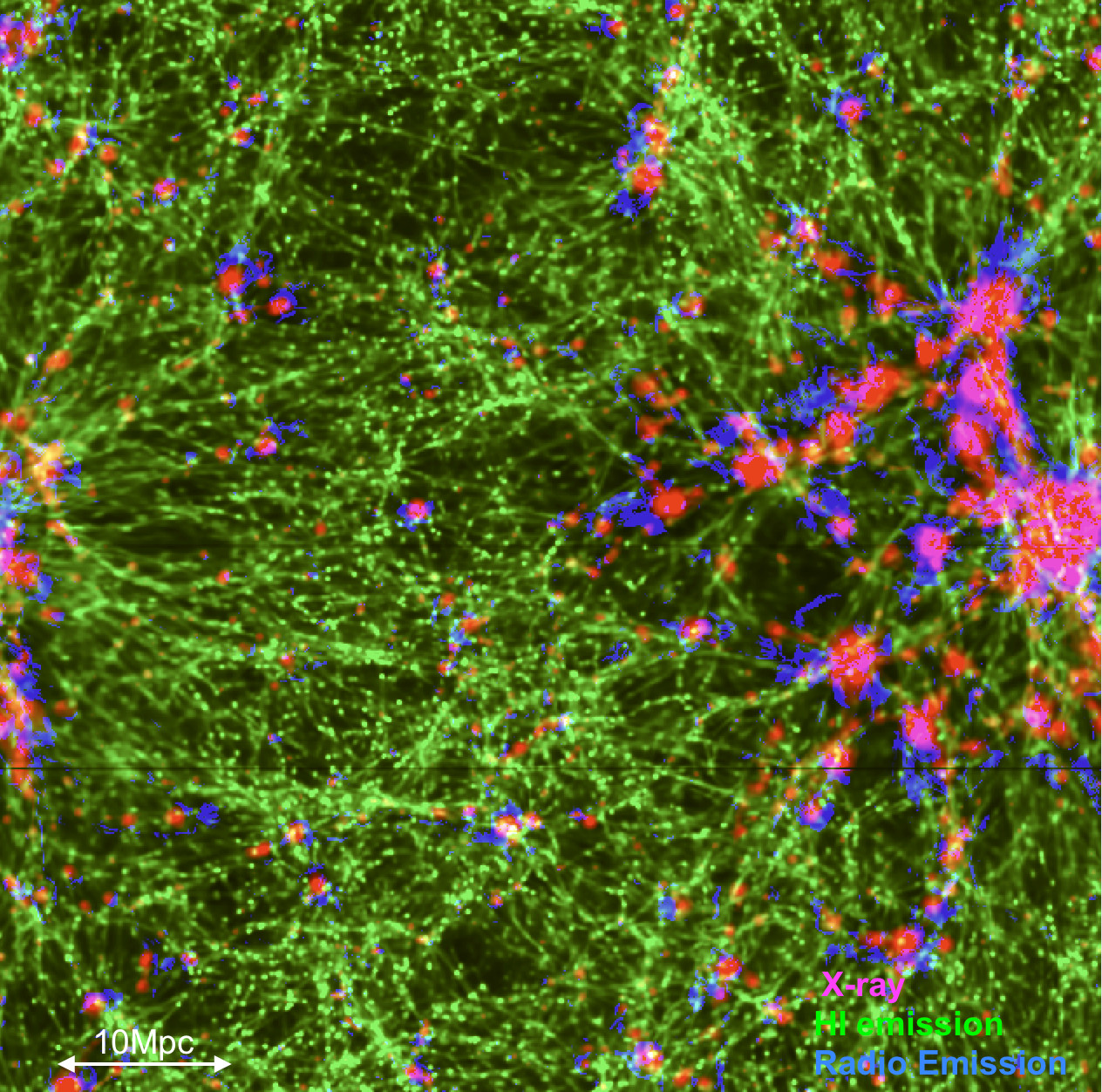}
\caption{RGB rendering of the distribution of different observable quantities for the same volume and model of Fig.\ref{fig:rgb}: synchrotron radio emission (blue), HI emission  (green) and X-ray emission  (pink).}
\label{fig:rgb2}
\end{figure}

\subsection{Multi-wavelength synthetic observations and halo catalogues}
\label{sec:images}

We generated emission maps of our simulated volume at different wavelengths, including a number of emission/absorption channels:

\begin{itemize}
    \item  {\it X-ray emission}:  we assume for simplicity a single temperature and a single (constant) composition for every cell in the simulation, and we compute the emissivity, $\Lambda$, from the B-Astrophysical Plasma Emission Code (B-APEC)  {\footnote{https://heasarc.gsfc.nasa.gov/xanadu/xspec/manual/Models.html.}}, computing  continuum and line emission under the assumption of collisional equilibrium, as in \citet{vazza19}. The metallicity is also assumed to be constant in each cell,  $Z/Z_{\odot}=0.3$. For each energy band, we compute the cell's X-ray emissivity and integrate along the line of sight (LOS): 
    
    \begin{equation}
    \rm S_X(E_1,E_2) [erg/s] =  \int n_H n_e \Lambda(T,Z) ~ dV 
    \end{equation}
    
    where  $n_H$ and $n_e$ are the number density of hydrogen and electrons (assuming a primordial composition) respectively, $T$ is the gas temperature and $dV$ is the constant volume of our cells {\footnote {Recently, \citet{2019MNRAS.482.4972K} included the contribution from the resonantly scattered cosmic X-ray background to the line emission for the warm-hot intergalactic medium in filaments, showing that this can increase its emissivity by a factor $\sim 30$.  However, the boost is limited to the gas at $T \leq 10^6 ~\rm K$, which gives little contribution to the [0.3-2.0 keV] band considered here.}. In this work we consider a broad energy range covering the soft X-ray spectrum (from 0.3 to 2.0 keV) as this was shown to yield the highest chances of detection with incoming X-ray telescopes \citep[e.g.][]{vazza19,2019arXiv190801778S}}.
    
    \item  {\it synchrotron radio emission}:  we compute the emission from relativistic electrons assuming they are accelerated by diffusive shock acceleration (DSA, e.g. \citealt{ka12} and references therein), accelerating a small fraction of thermal electrons swept by shocks up to relativistic energies ($\gamma \geq 10^3-10^4$).  We identify shocks in our simulations in post-processing, with a velocity-based approach \citep[][]{va09shocks}, and we compute the radio emission from electrons accelerated in the shock downstream following  \citet{hb07}. The radio emission is calculated in  post-shock cells as the convolution of the several power-law distributions of electrons that overlap in the downstream cooling region, to which we assign an integrated radio spectrum:

\begin{equation}
P(\nu) \rm [\frac{erg}{s \cdot Hz}]=6.4 \cdot 10^{34} \int  \frac{S ~n_e \xi_e(\mathcal{M}) ~T^{3/2}}{\nu^{s/2}} \cdot \frac {B^{1+s/2}}{B_{\rm CMB}^2+B^2} dV,
\label{eq:hb}
\end{equation}

where $S$ is the shock surface, $\xi_e(\mathcal{M})$ is the acceleration efficiency of electrons as a function of Mach number \citep[see][for details]{va17cqg}, $\nu$ is the observing frequency, $B$ is the magnetic field strength in the post-shock cell and $B_{\rm CMB}$ is the magnetic field-equivalent to the Cosmic Microwave Background energy density ($B_{\rm CMB}=3.2 \rm ~\mu G (1+z)^2$). 
 Our model does not account for radio galaxies, which are an important contributor to the radio emission from the cosmic web. However, our masking procedure (see Section 3.1) makes the presence of radiogalaxies in the simulation irrelevant, as the cross-correlations in such case are computed after removing their putative location from the sky model.  Moreover, we do not include the contribution from the additional diffuse radio emission which may be produced by secondary electrons \citep[][]{de00} and/or turbulent re-acceleration \citep[][]{2009A&A...507..661B}. Both these scenarios have been proposed for the origin of  ``radio halos" \citep[e.g. see][for recent reviews]{bj14,2019SSRv..215...16V}, but their contribution should be largely sub-dominant compared to the radio emission from cosmic shocks starting from the periphery of halos \citep[e.g.][]{va15survey}.
    
\item {\it Faraday Rotation (RM)}: we define for each beam of cells along the LOS in each map the Faraday Rotation experienced by linearly polarised radio emission as
    \begin{equation}
  \rm RM \ \rm[rad/m^2]=812 \int \frac{B_{\rm ||}}{\rm [\mu G]} \cdot \frac{n_e}{\rm [cm^3]} \frac{dl}{[kpc]}\frac{1}{1+z},
\end{equation}
where $\rm ||$ denotes the component of the magnetic field parallel to the LOS, $z$ is the redshift of each cell,  $n_e$ is the physical electron density of cells, assuming a primordial chemical composition ($\mu=0.59$) of gas matter everywhere in the volume \citep[e.g.][]{va17cqg}. 

\item {\it Sunyaev-Zeldovich effect (SZ)}: we compute the SZ  signal at 21 cm of the specific intensity of the CMB at the frequency $\nu$, assuming a small optical depth everywhere, as 

\begin{equation}
\Delta I_{\rm SZ}(\nu) [\rm Jy/sr^2] = \frac{4 k_{\rm b}^2 \sigma_T T_{\rm CMB} }{m_e c^2}  \left (\frac{\nu}{c} \right )^2  \int  \rm \frac{dl}{\rm [cm]}\, \frac{n_e}{\rm [cm^3]} \frac{T}{[K]} 
\end{equation}
where $T_{\rm CMB}$ is the CMB temperature, which is appropriate for the  Rayleigh-jeans part of the CMB spectrum \citep[e.g.][]{1999PhR...310...97B}, $\sigma_T$ is the Thomson cross section, $k_b$ is the Boltzmann constant, $m_e$ is the electron mass and $c$ is the speed of light. 

\item  {\it Neutral Hydrogen radio emission}: 
we estimate the spin temperature of HI and its related signal following \citet[][]{2017PASJ...69...73H}. In summary the spin temperature is computed assuming three physical processes : a) the excitation and de-excitation by the CMB photons; b) collisions with electrons and other atoms; c) interactions with
background Lyman-$\alpha$ photons \citep{1959ApJ...129..536F}:
\begin{equation}
	T_{\rm HI}\rm [K]=\frac{1+x_{\rm c}+x_{\alpha}}{T_{\rm{CMB}}^{-1}+x_{\rm c}T^{-1}+x_{\alpha}
	T_{\alpha}	^{-1}} , \label{spin}
\end{equation}

where $x_{\rm c}$ and $x_{\alpha}$ are the coupling coefficients for the collisional process and the interaction with Ly$\alpha$ photons, while  $T_{\alpha}$ is the colour temperature in the vicinity of the Ly-$\alpha$ frequency. Formulas for $x_c$, $x_\alpha$, $T_\alpha$ and for the Lyman-$\alpha$ background mean intensity $J_{\alpha}$,  are given in \citet{2017PASJ...69...73H}. 

While in CSFBH2 model the HI abundance is computed in a self-consistent way by {\enzo} chemistry modules, in the non-radiative runs we simply assume a fixed reference value of $10^{-6}$ for the fraction of the gas density, value that is commonly found in simulations \citep[e.g.][]{2009A&A...504...15P} at gas temperatures $T \leq 10^{5-6} \rm K$. In principle, the contribution from HI can be computed even in a simple non-radiative simulation, assuming that the thermal and ionization evolution of gas in filaments is fully determined by the combined effect of the background UV radiation and of the Hubble expansion \citep[e.g.][]{2018ApJ...866..135V}. However, the shock heating by large scale shocks and, possibly, the feedback from active galaxies, both included in our simulations, are expected to affect the properties of filaments in HI. Neglecting these effects leads to underestimating the temperature in filaments and this grossly overestimates the neutral hydrogen fraction. Therefore, only our CSFBH2 model includes the necessary physics to provide a robust estimate of the HI temperature, while the other two runs are only presented for completeness, but are not realistic enough.\\

Galaxy and galaxy clusters/groups have been extracted as Dark Matter (DM) halos from the simulated data using the following procedure. First, we identify the local maxima in the DM mass density field above a given threshold, which is set according to the typical number of objects expected from a given observation. For instance, setting the threshold $\rho_{th} = 2 \cdot 10^{-30} \rm g/cm^3$, our procedure identifies $N_g \approx 220,000$ halos in the $85^3$ $\rm Mpc^3$ volume, corresponding to a density of $\sim 0.35$ galaxy per comoving $\rm Mpc^3$ and a projected galaxy density of $\sim 3520$ galaxies per square degree, consistent with the expected number count of galaxies at low redshift by Euclid, which is $\sim 3 \cdot 10^{3}-10^{4}$ galaxies per square degree \citep[e.g.][]{2012MNRAS.427.3134B}. A halo is then reconstructed around each peak,  with a spherical overdensity algorithm which computes the radial mass profile of halos within concentric shells. Increasingly larger spheres are built around the peak until the internal average mass overdensity is equal to N. The corresponding radius, $R_{\rm N}$, defines the size of the halo. In our case, we have produced the halo catalogues for $R_{\rm 100}$ and $R_{\rm 200}$. The case $R_{\rm 500}$ has also been calculated, being more in line with real X-ray observations at the scale of groups of galaxies \citep[][]{2011A&A...535A...4R,2011A&A...535A.105E} discussed in Section \ref{sec:validation}.
Small objects, typically with masses $M_{\rm 100} \leq  10^{10} \rm M_{\odot}$, can have a diameter close to the resolution of the simulation. In that case, we assign to the galaxy the radius corresponding to single computational cell.


\end{itemize}

An RGB rendering of the projected X-ray emission, radio emission and HI emission in our CSFBH2 model at $z=0.045$ is shown in Fig.\ref{fig:rgb2}, well illustrating once more how the different emission proxies considered in this work have different level of clustering with the underlying large-scale distribution of the cosmic web. 

For the sake of simplicity we neglect in all cases the contribution from the Galactic Foreground, as well as the intrinsic contribution from galaxies (e.g. internal Faraday Rotation or X-ray/radio emission from active galactic nuclei) as they cannot be properly resolved in our cosmological simulations.

As a reference, our simulated volume has been located at $z \approx 0.045$ ($d_L \approx 200$ Mpc). Images consist of $1024\times1024$ pixels, meaning that each simulated sky model has a pixel resolution of $\approx 90 ~\rm "/pixel$ for the assumed cosmology.  

\subsection{Validation of models}
\label{sec:validation}

As a preliminary step, we have validated the models adopted in this work in order assess their reliability for the objectives of our study. Additional tests (for the larger set of the Chronos++ suite of simulations) can be found in \citet{va17cqg}. 

\begin{figure}
\includegraphics[width=0.48\textwidth]{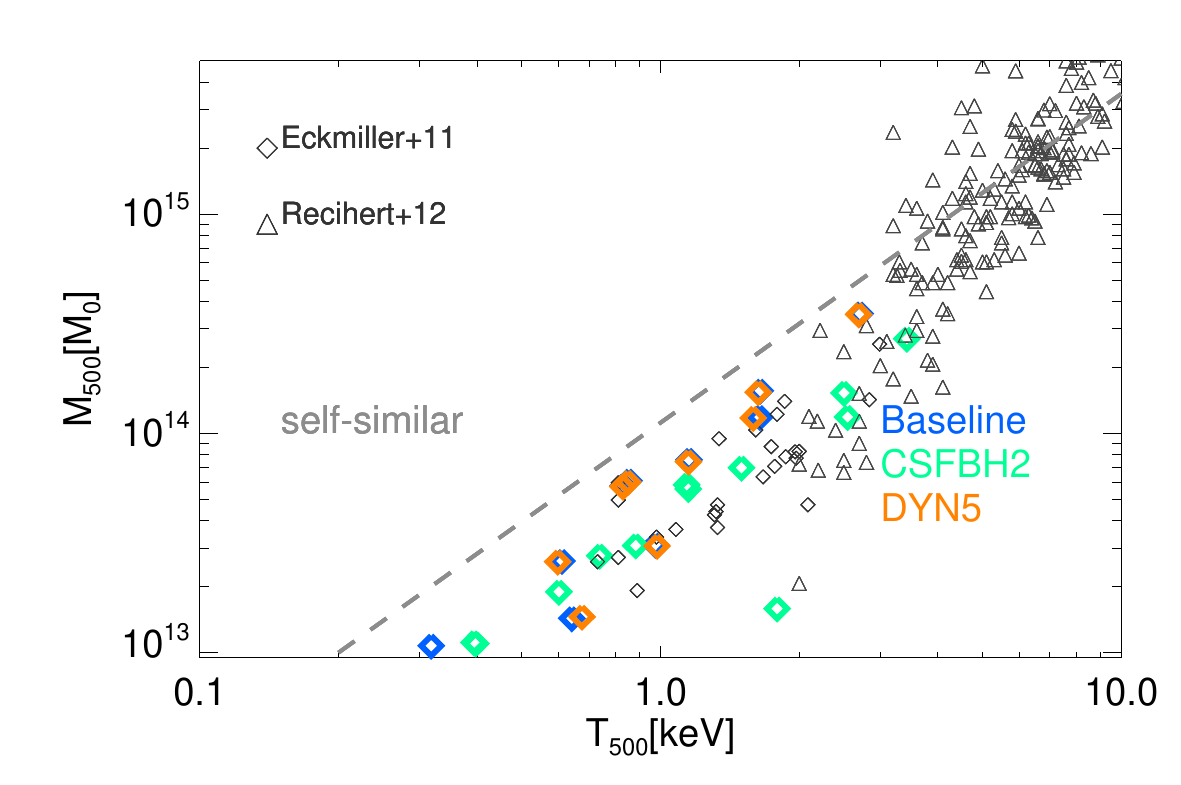}
\caption{Gas temperature-Total Mass relation  at $z=0$ within $R_{\rm 500}$ for clusters in our three models. The dashed line shows the self-similar scaling relation and the additional grey points show the X-ray observations by \citet{2011A&A...535A.105E} and \citet{2011A&A...535A...4R}, for comparison. }
\label{fig:scaling}
\end{figure}

\begin{figure}
\includegraphics[width=0.48\textwidth]{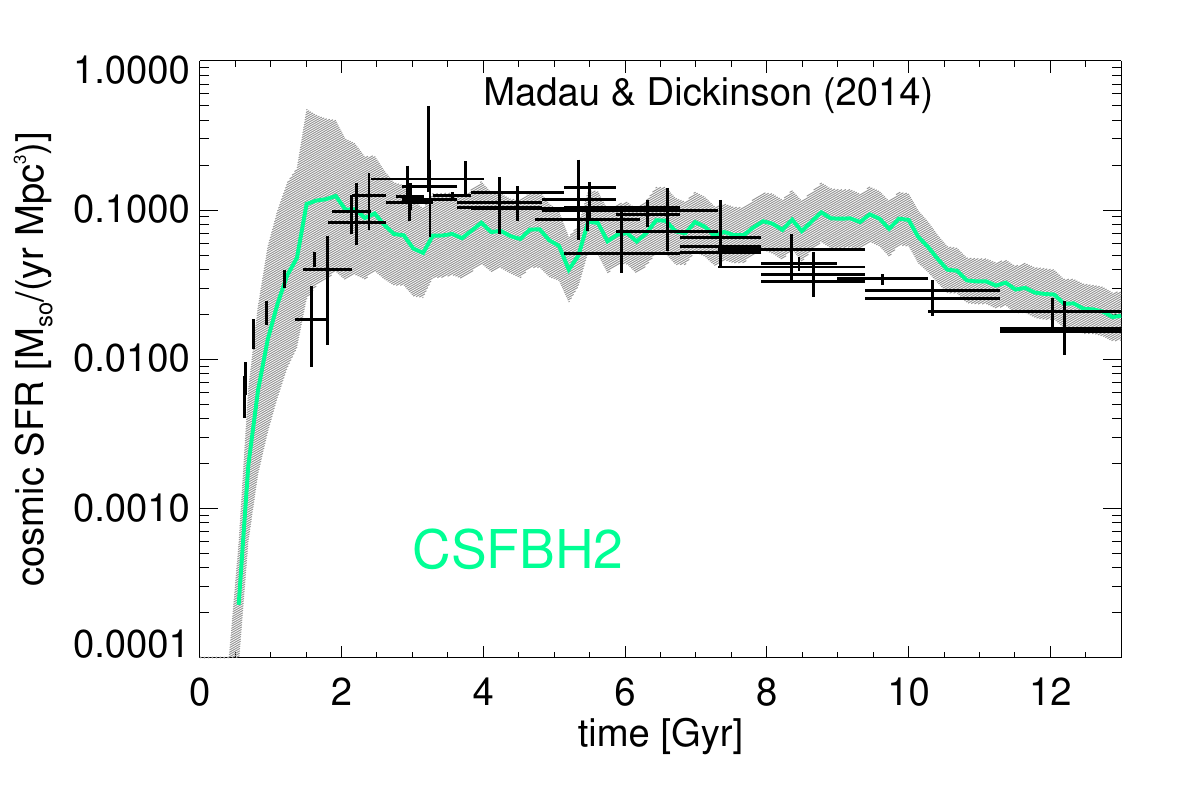}
\caption{Simulated cosmic star formation history for our CSFBH2 run (coloured lines with $\pm 3 \sigma$ variance) against the observed cosmic star formation history from the collection of observations in  \citet{2014ARA&A..52..415M} (black points with errorbars).}
\label{fig:sfr}
\end{figure}

\begin{figure}
\includegraphics[width=0.48\textwidth]{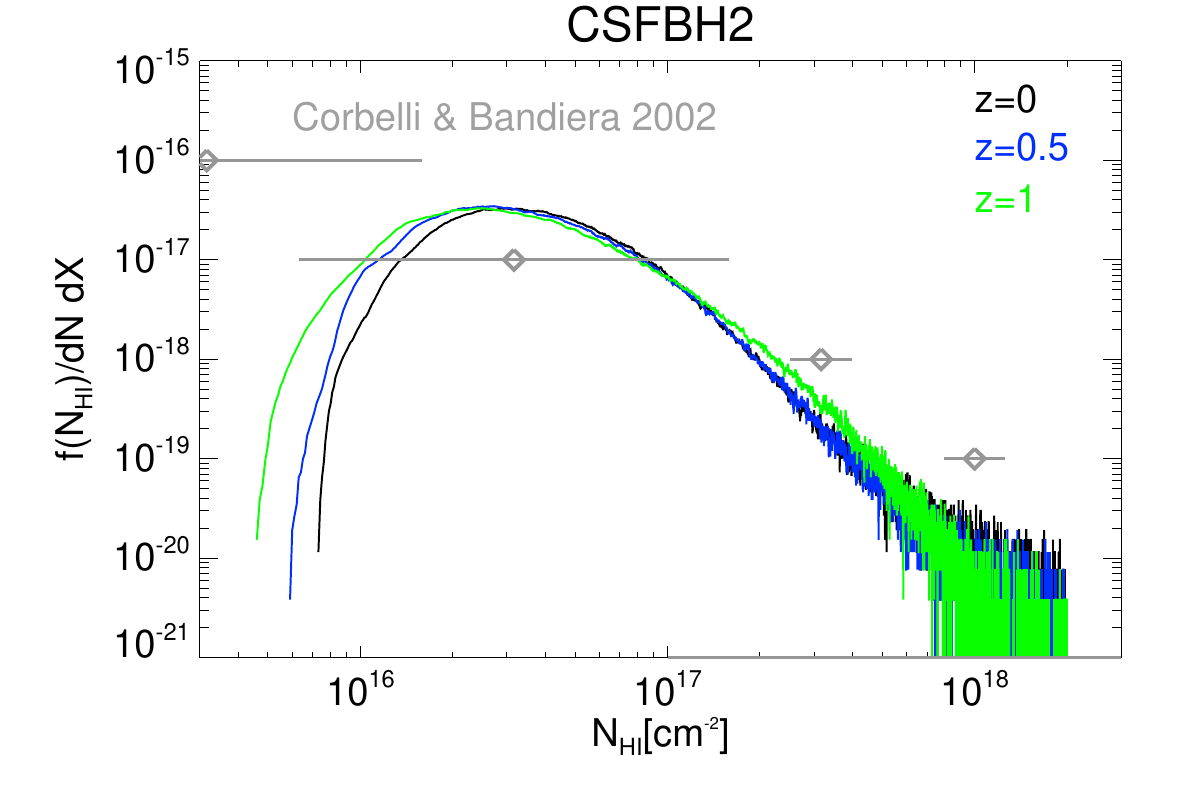}
\caption{Differential distribution of HI column density at three epochs in our CSFBH2 run. The additional gray points with errorbars are derived from \citet[][]{2002ApJ...567..712C}.}
\label{fig:dNHI}
\end{figure}

\begin{figure}
\includegraphics[width=0.45\textwidth]{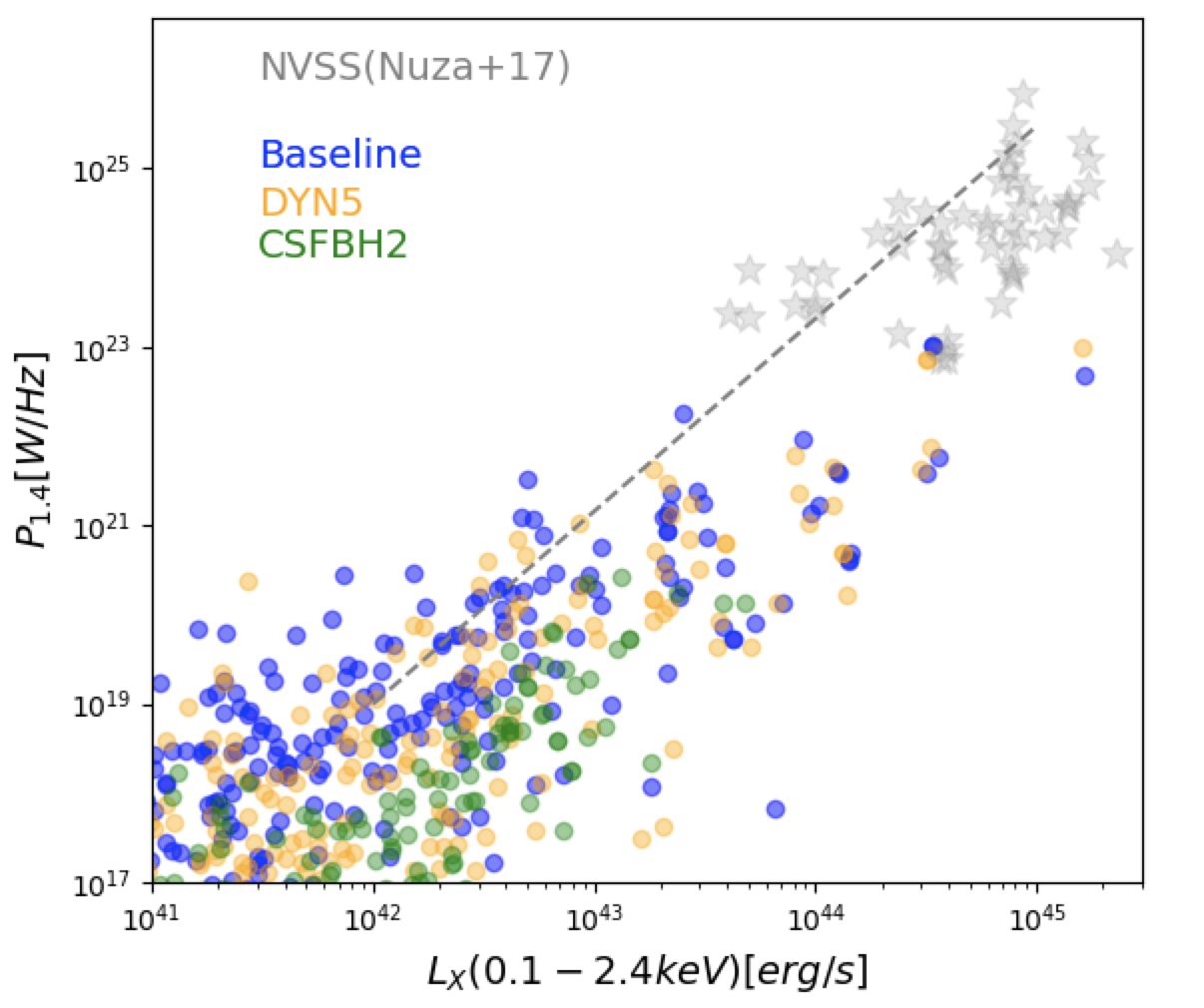}
\caption{Simulated vs observed scaling relation between the X-ray luminosity in the [0.1-2.4] keV band and the total radio power at 1.4 GHz from radio relics. The observational data (stars) are taken from \citet[][]{2017MNRAS.470..240N}, while the black dashed line is derived from the best-fit relation by \citet{fdg14}.}
\label{fig:relics}
\end{figure}

The Mass-Temperature scaling relations of groups and clusters of galaxies in our simulations is presented in Figure \ref{fig:scaling}. The clusters/groups are identified as spherical halos at $R_{\rm 500}$ (see Section \ref{sec:images}). The results are similar to what already discussed in \citet{va17cqg} for the larger set of simulations of our Chronos++ suite: while the non radiative runs (baseline and DYN5) strictly follow, as expected, the self similar $T \propto M^{2/3}$ relation, the combined effects of radiative cooling, star formation and feedback from black holes in the CSFBH2 run  steepens the scaling relation within $R_{\rm 500}$, more in line with real X-ray observations at the scale of groups of galaxies  \citep[][]{2011A&A...535A...4R,2011A&A...535A.105E}. This is an effect of the increased temperature within $R_{\rm 500}$ of $M_{\rm 500} \leq 10^{14}M_{\odot}$ systems, in which 
the total feedback energy is of the same order of the gas potential energy. This ensures that, in general, the large-scale distribution of thermal gas in our simulations is realistic enough compared to the expected properties of galaxy clusters. In addition, the introduction of AGN feedback in run CSFBH2 modifies our lowest mass systems in a way compatible with observational data, for realistic feedback parameters. 

Figure \ref{fig:sfr} shows the cosmic star formation (SFR) history for run CSFBH2, compared to the survey of infrared and ultraviolet observations from \citet{2014ARA&A..52..415M}. The match is reasonably good at all epochs/redshift, indicating that overall our ad-hoc prescription for star formation and feedback performs well in converting the gas cooling within halos into star forming particles, as well as that the amount and duty cycle of feedback in our halos is fairly compatible with observations (see \citealt{va17cqg} for more details).

In Figure \ref{fig:dNHI}, we show the distribution of HI column density for three  different epochs ($z=0.0$, $z=0.5$ and $z=1.0)$ in the CSFBH2 model, compared to observations \citep[e.g.][]{2002ApJ...567..712C,2017ApJ...849..106S}. At all epochs, the agreement is far from being satisfactory, with a lack of HI absorbers both for $N_H \leq 5 \cdot 10^{15} \rm ~cm^2$ and $N_H \geq 5 \cdot 10^{18} \rm ~cm^2$. 
This suggests that the emergence of neutral hydrogen in our model is undermined by the insufficient spatial resolution, which is a key factor for the formation of HI in the circumgalactic medium, as well as at low column densities \citep[e.g.][]{2018arXiv181112410H}. This is caveat to be considered in interpreting our results in the remainder of the paper. 

Figure \ref{fig:relics} gives the scaling between the integrated synchrotron emission at 1.4 GHz and the X-ray luminosity in the [0.1-2.4 keV] band of host clusters within, in this case within $R_{\rm 200}$ (i.e. $200$ time the critical cosmic matter density) to compare with the observed statistics of "radio relics" in galaxy clusters \citep[e.g.][]{2017MNRAS.470..240N}, which are believed to represent the tip of the iceberg of the distribution of radio emission in the cosmic web. Despite the lack of massive halos, due to the limited volume simulated here,  we can compare with observations by extrapolating to lower masses/X-ray luminosity the  $(P_{\rm 1.4},L_{\rm X})$ scaling found by \citet[][]{fdg14}, as shown in the figure. 
Albeit with a large scatter, the relic radio emission from our population of clusters is reasonably consistent with the observed distribution of radio relics, that in turn suggests that the synchrotron model adopted here (see Eq.2) is plausible for all models as it does not violate available radio constraints. \\

\begin{figure}
\includegraphics[width=0.5\textwidth]{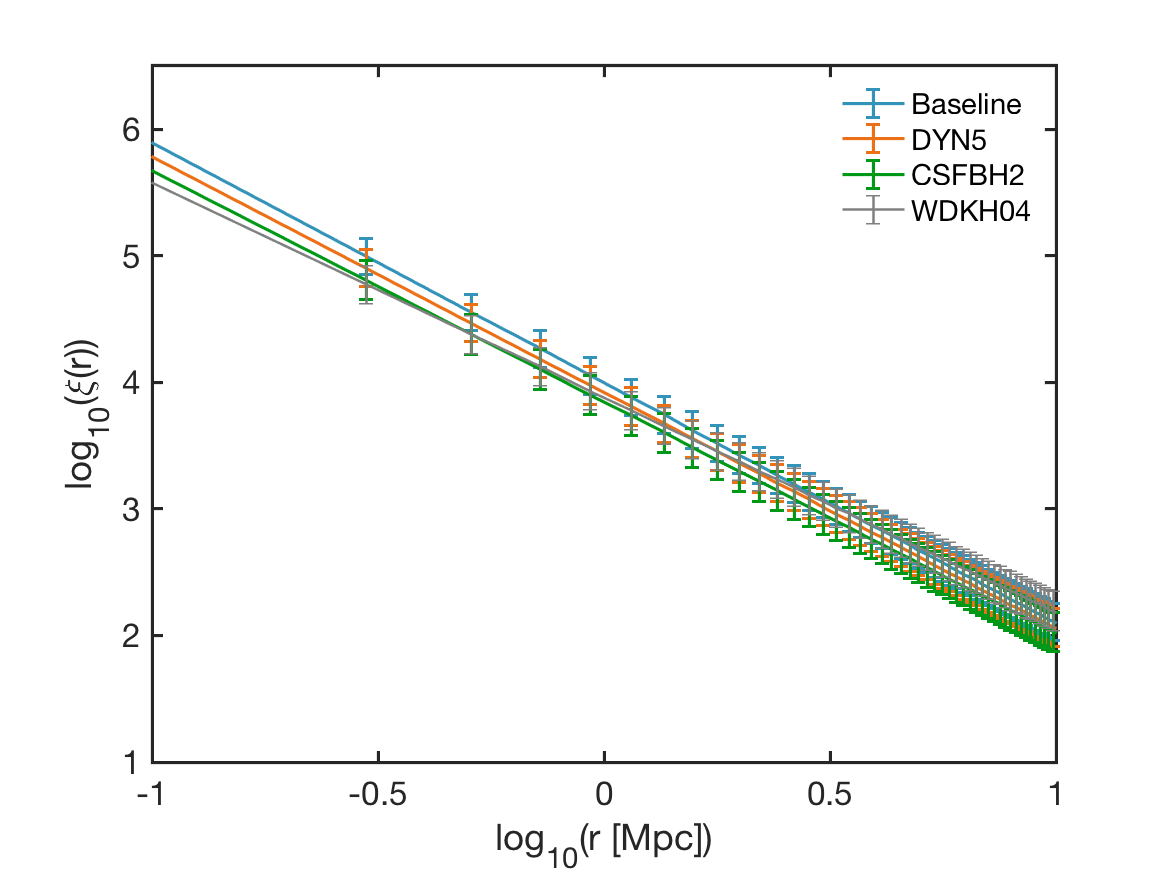}
\caption{Two-points correlation function in the range 0.1 to 10 Mpc, in arbitrary units, of the halos extracted from the three runs (coloured lines), compared with that from \citet{2004ApJ...601....1W} (gray line). 3-$\sigma$ error bars are shown.}
\label{fig:correlation}
\end{figure}

Despite the aforementioned difficulty in resolving galaxy formation physics with our runs (Section \ref{sec:simulations}) and in properly identifying small objects (Section \ref{sec:images}), the statistical large-scale clustering properties of the halos within filaments are well resolved and described by our approach, as shown by Figure \ref{fig:correlation}, that presents their two-points correlation function. For all models, the function resulted to follow a power law with exponent $\gamma = -1.899\pm0.0657$ for the baseline model, $\gamma = -1.867\pm0.067$ for DYN5 and $\gamma = -1.828\pm0.071$ for the CSFBH2 model, in the range 0.1 to 10 Mpc, consistent, although slightly steeper (given the reduced small scale power due to the limited resolution) with that expected for galaxies in the standard $\Lambda$CDM model \citep[][]{2004ApJ...601....1W}. This was shown also in \citet{gh16}, where the same procedure was adopted to compare the resulting halo catalogues with the GAMA data \citep[e.g.][]{alp14a}.
The normalisation (in arbitrary units) of the various correlation functions differs only slightly. The similarity of the different curves points out how the clustering properties of the halos are only mildly influenced by the different physical setup characterising the various simulations.\\

In summary, the physical models used in this work are realistic enough to present a first systematic study of the cross-correlation between different observable signatures of diffuse gas and magnetic fields in the cosmic web.

\section{Cross-correlation }
\label{sec:cross}

Cross-correlation analysis is commonly used in signal processing to measure the similarity of two signals as a function of the displacement of one relative to the other. For 2D $N\times M$ pixels images $A$ and $B$, the normalised correlation matrix $C$ is defined as:
\begin{equation}
    C(k,l) = {1\over NM}\sum_{j=0}^{N-1} \sum_{i=0}^{M-1} {(A(i,j)-\bar A)(B(i+k,j+l)-\bar B) \over 
               \sigma_A \sigma_B},
\end{equation}
where $\bar A$ and $\bar B$ are the mean values of the two images and $\sigma_A$ and $\sigma_B$ are their standard deviation. The indices of the correlation matrix represents the shift (displacement) of the two images. The correlation is normalised by the standard deviation of each quantity in order to allow for a direct comparison of quantities that can differ by many orders of magnitude (as in our case). The normalised cross-correlation function $C_r$ is calculated as the average of $C(k,l)$ over elements having the same radial separation $r = (k^2 + l^2)^{1/2}$.  This 1D averaging assumes radial symmetry in the 2D function, which is guaranteed due to the cosmic isotropy condition assumed in the simulations. The $C_r$ function takes values between -1 and 1, the latter representing perfect linear correlation between quantities ($A\propto B$). The value -1, represents perfect linear anti-correlation. 

\begin{figure}
\includegraphics[width=0.45\textwidth]{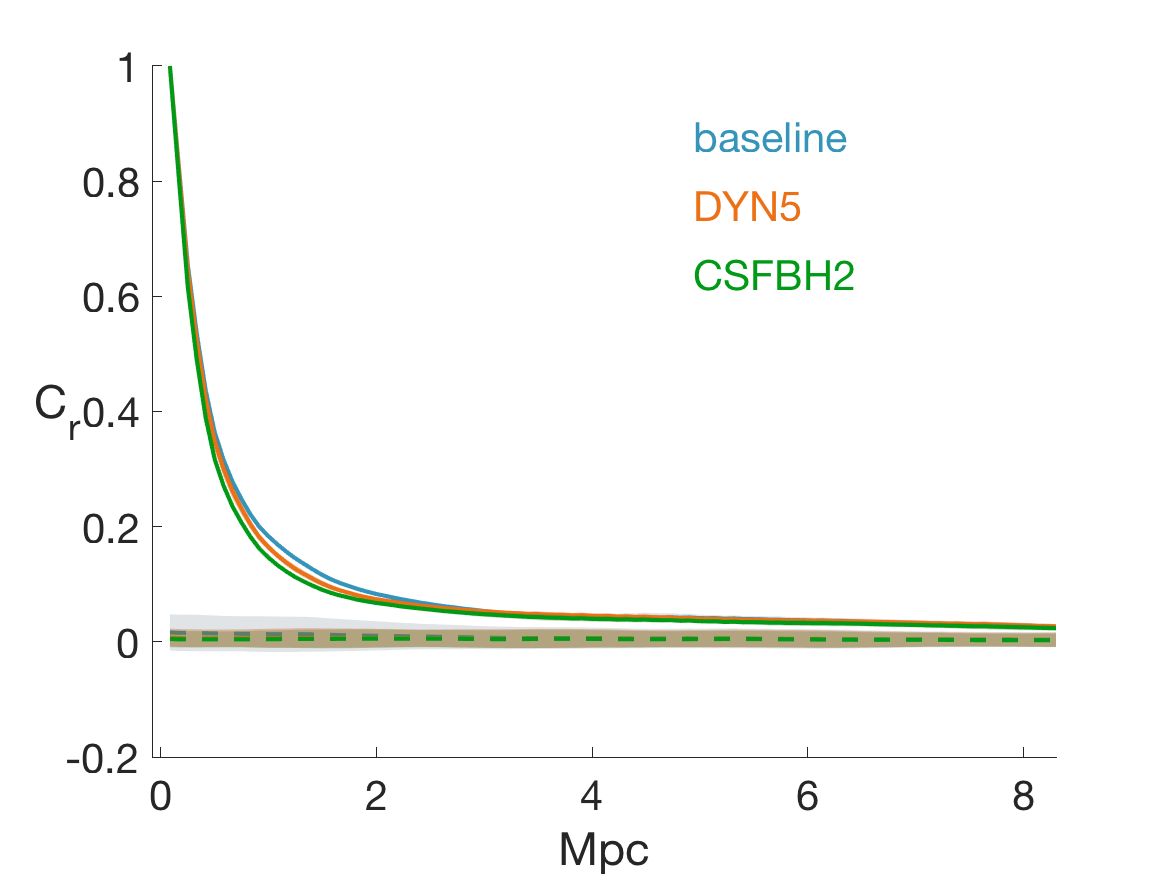}
\caption{Autocorrelation function of the Dark Matter mass density distribution for the three models. Solid lines are calculated cross-correlating each image with itself and averaging over the three orthogonal projections. Dashed lines represent the null reference baselines, calculated cross-correlating each image with one of the remaining two projections and averaging the the results.}
\label{fig:auto}
\end{figure}

In the case of our maps, the significance of the cross-correlation is evaluated against the case of null correlation, for which $C_r$ is computed between $A$ and $B$ from two different projections. As an example, Figure \ref{fig:auto} shows the autocorrelation function of the Dark Matter mass density distribution for the three models, calculated cross-correlating each image with itself, compared to the same quantity but calculated correlating each image with one of the remaining two projections. Such comparison gives an indication of the shape of the signal expected from perfectly correlated images, equal to one at zero displacement and then monotonically decreasing with increasing shifts, compared to that of perfectly uncorrelated ones, which is randomly fluctuating around zero at all displacements.

\subsection{Masking of halos}
\label{sec:masking}

In order to separate correlation signals due to gas within halos and outside them, we calculate the cross-correlation both on full maps, and on ``masked" ones, excising the information coming from the halos, and focusing on the contribution from the cosmic web only. Masking allows also to exclude those regions where the limited resolution of our models may have a major impact on the simulated formation of galaxies and on their impact on our observables. 

\begin{figure*}
\includegraphics[width=0.9\textwidth,height=0.185\textheight]{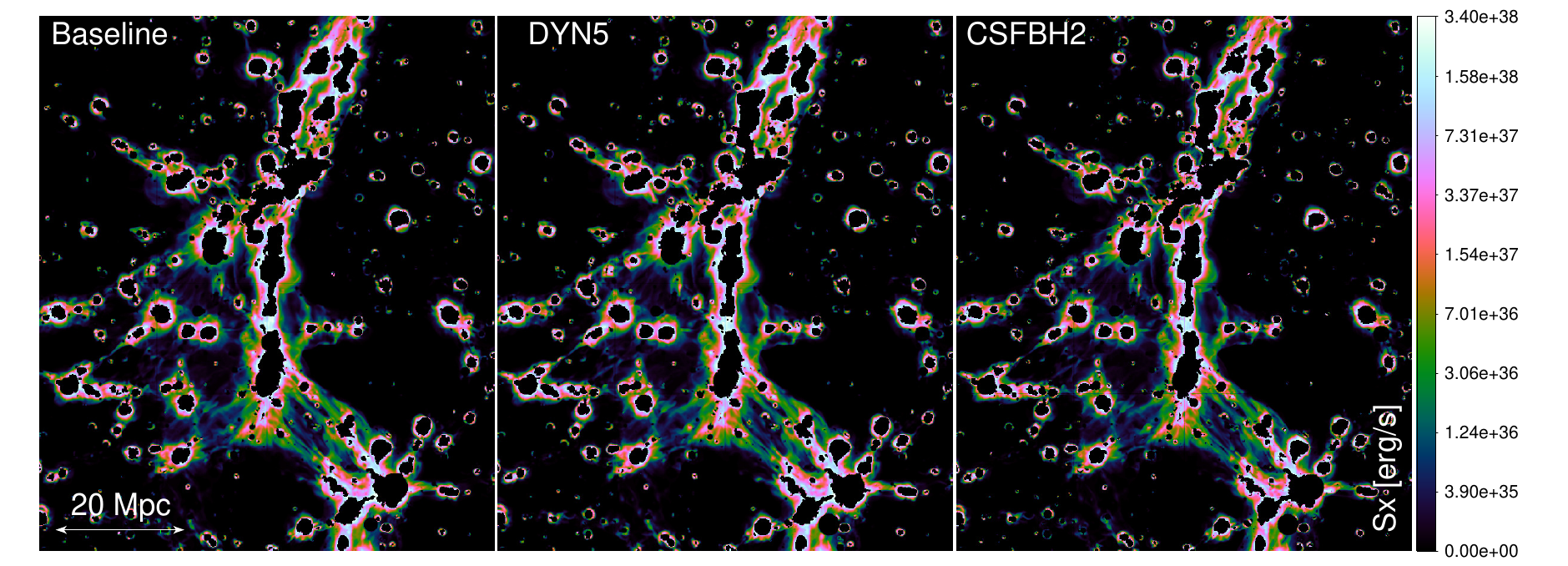}
\includegraphics[width=0.9\textwidth,height=0.185\textheight]{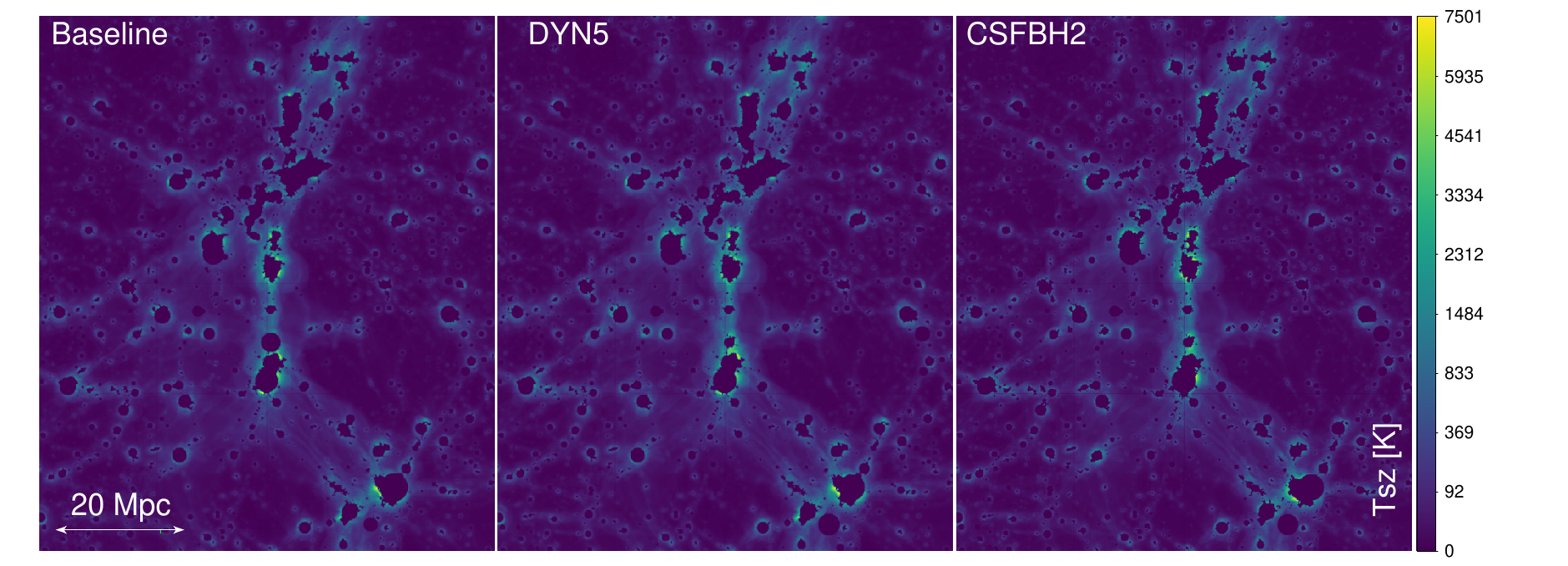}
\includegraphics[width=0.9\textwidth,height=0.185\textheight]{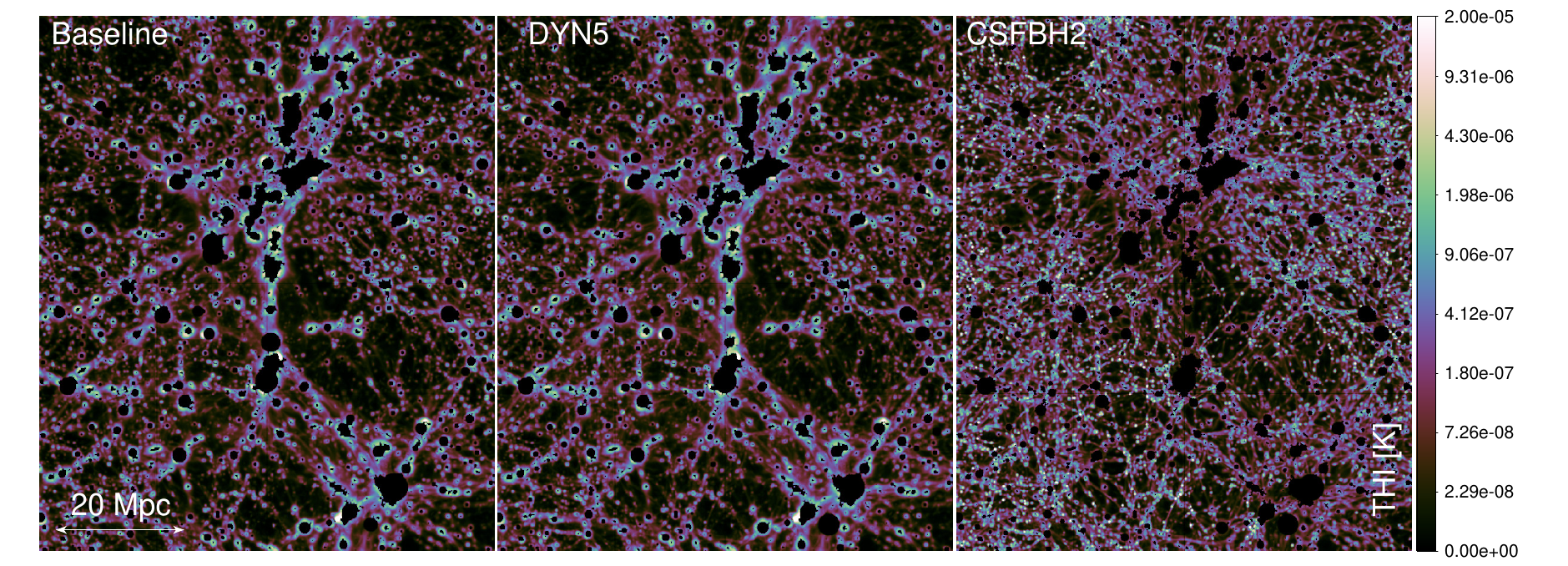}
\includegraphics[width=0.9\textwidth,height=0.185\textheight]{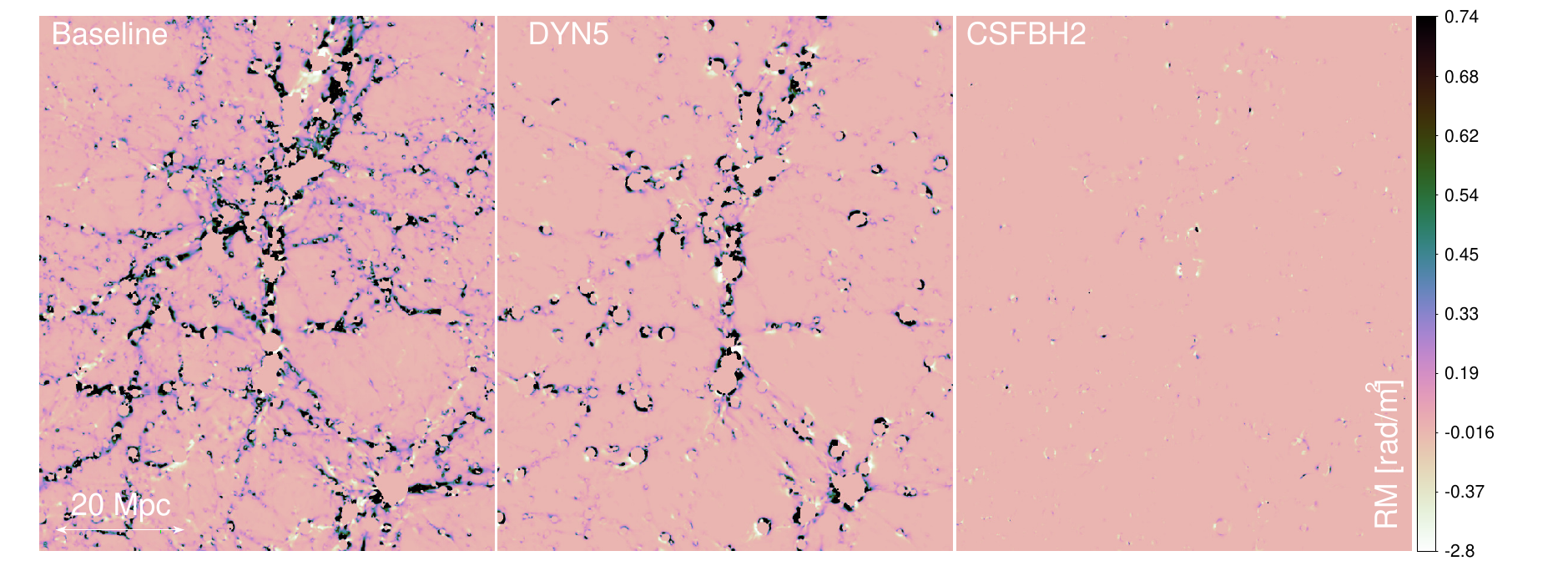}
\includegraphics[width=0.9\textwidth,height=0.185\textheight]{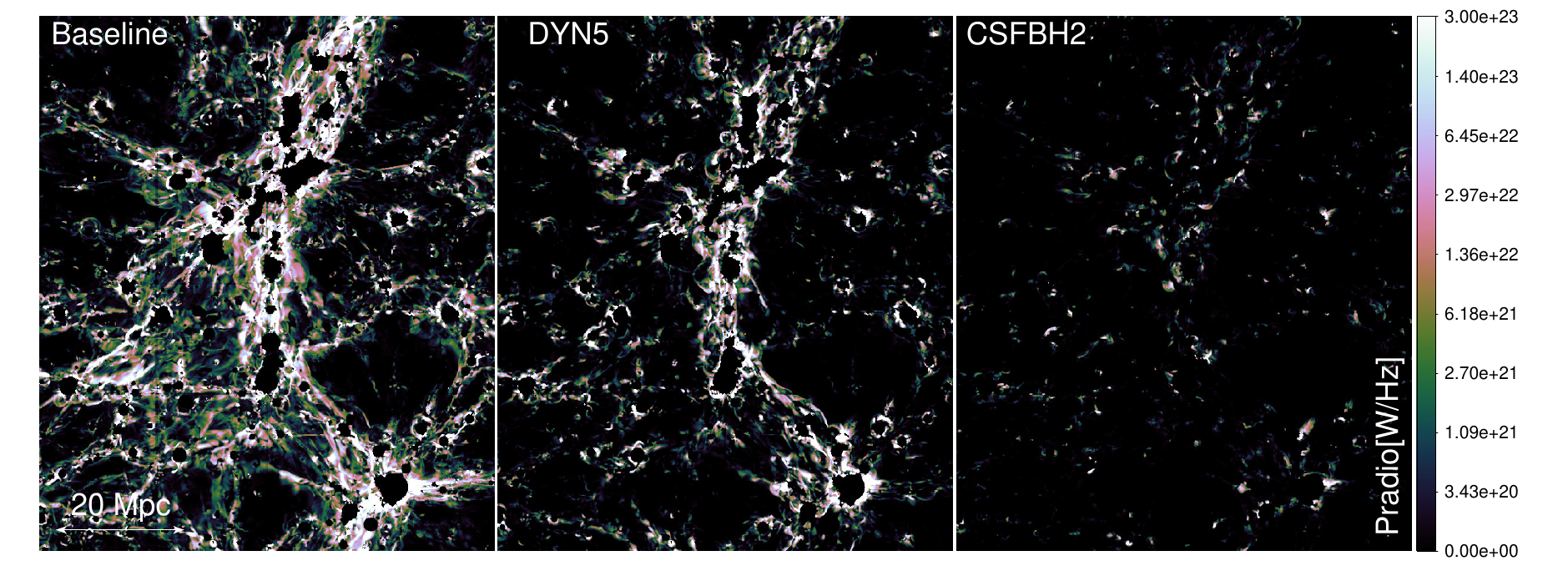}
\caption{Example of projected maps of various observables: from top to bottom: X-ray emission in the [0.3-2]keV band, Sunyaev-Zeldovich signal computed at $200 \rm GHz~$, HI brightness temperature at 1.4 GHz, Faraday Rotation measure and synchrotron radio emission at 200 MHz, for our three models and for a $85^3 \rm ~Mpc^3$ volume located at $z=0.045$. $R_{\rm 100}$ masking is applied.}
\label{fig:maps}
\end{figure*}

Masking is performed projecting the halo catalogues on the maps and setting to $-1$ all the pixels within circles centred on each halo centre and radius $R_{\rm N}$. In calculating the cross-correlation, all pixels with value $-1$ can then just be discarded. We have tested two different masks, corresponding to the catalogues $R_{\rm 100}$ and $R_{\rm 200}$ respectively (see Section \ref{sec:images}). We anticipate that the results obtained adopting the two different masking radii do not show meaningful differences in all cases, therefore in the rest of the paper we discuss the $R_{\rm 100}$ case only. Examples of the $R_{\rm 100}$ masked maps are shown in Figure \ref{fig:maps}.

\section{Results}
\label{sec:analysis}

In this section we present the results obtained calculating the cross-correlation between the quantities introduced in section \ref{sec:images}, for the different models. 

\begin{figure}
\includegraphics[width=0.45\textwidth]{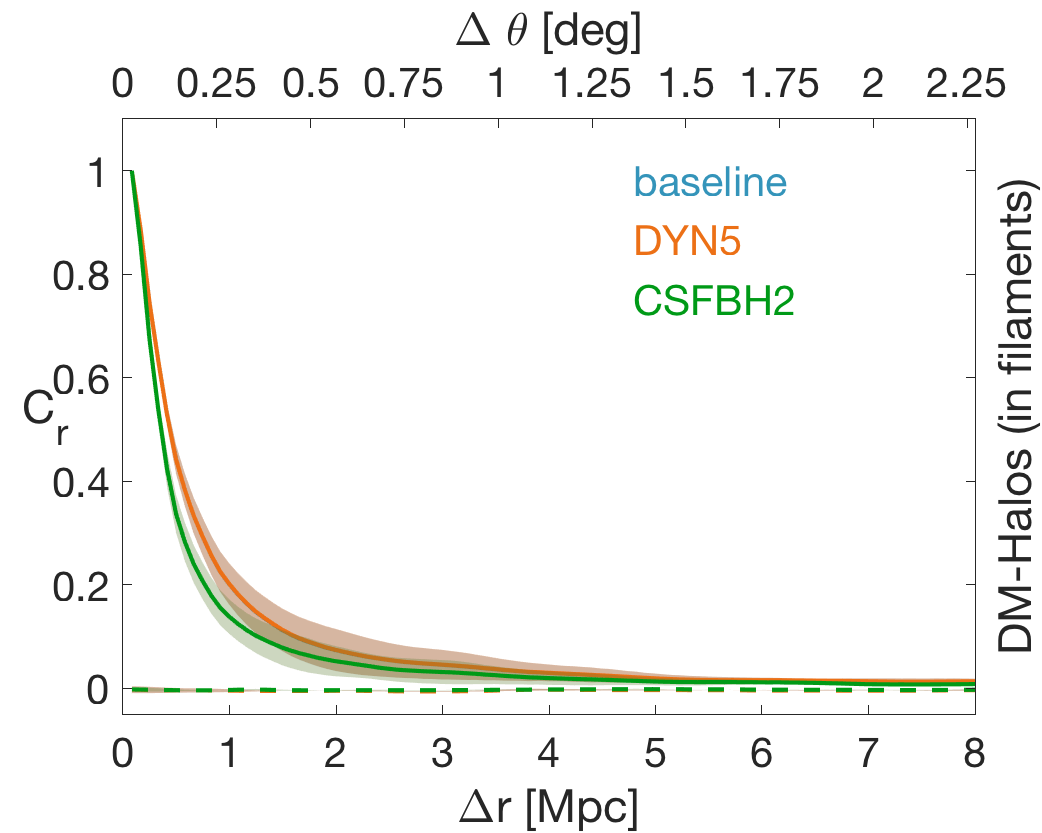}
\caption{Cross-correlation between the DM mass and the halo distributions. Halos are identified as local maxima of the DM density field. In the base horizontal axis, the scale is in Mpc, in the upper horizontal axis the scale is in degrees.} 
\label{fig:halos}
\end{figure}

Figure \ref{fig:halos} shows the cross-correlation between the halo maps and the DM mass density, showing the expected correspondence between halos and mass distributions. The baseline and the DYN5 models are indistinguishable, magnetic fields having no meaningful impact on the dark matter dynamics. For the CSFBH2 model the correlation drops faster than for the other two models with separation. In this case, in fact, cooling and energy feedback affects the gas dynamics, with a non negligible feedback on the overall mass distribution. The loss of correlation above 0.5 Mpc is due to the formation of more compact collapsed objects, due to the cooling, combined with the effect of AGN outflows, wiping out overdensities in affected regions. 
  
\subsection{Cross-correlation analysis of physical models}

We first study the cross-correlation of different quantities, without the inclusion of any observational noise or instrumental effect, focusing on the impact of the model variations of our set of simulations.

\subsubsection{Cross-correlation with the Dark Matter distribution}
\begin{figure}
\includegraphics[width=0.23\textwidth]{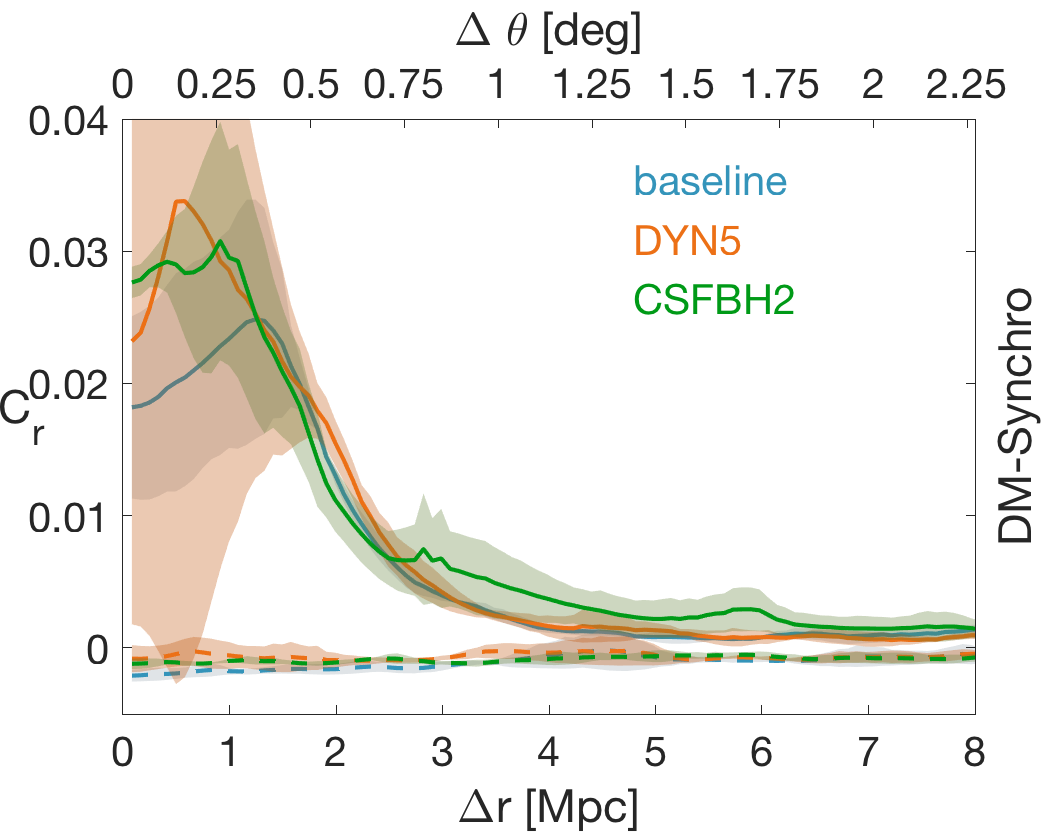}
\includegraphics[width=0.23\textwidth]{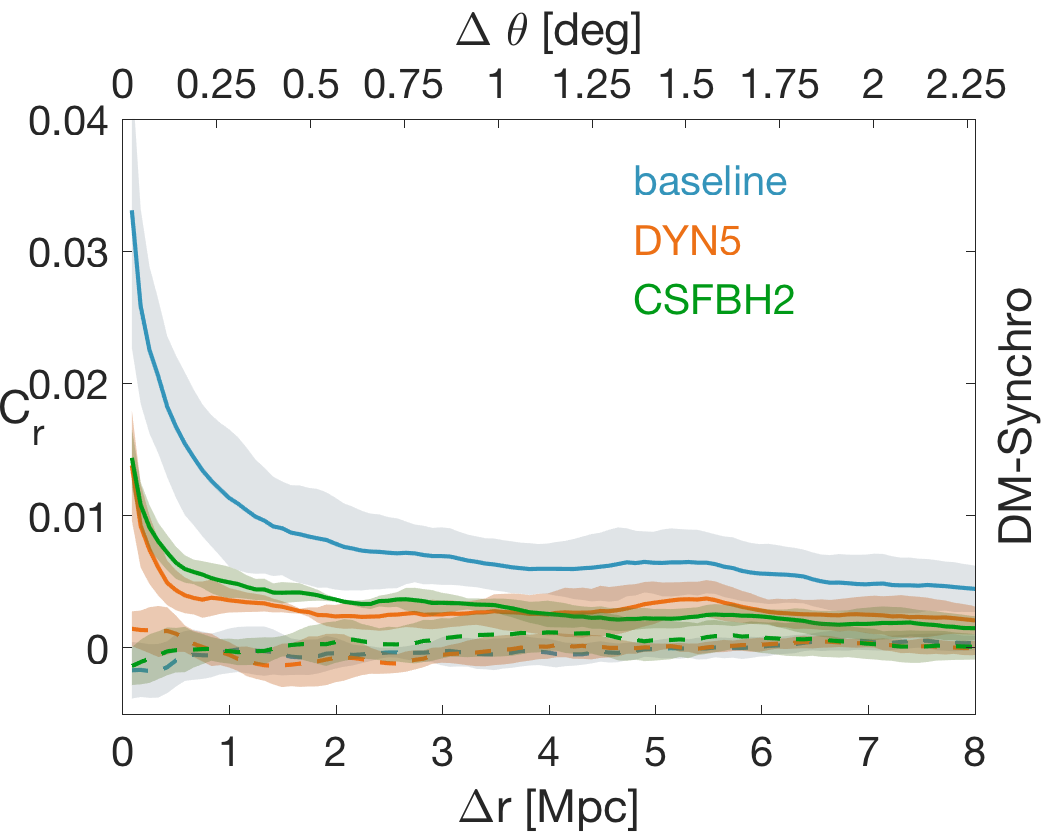}
\includegraphics[width=0.23\textwidth]{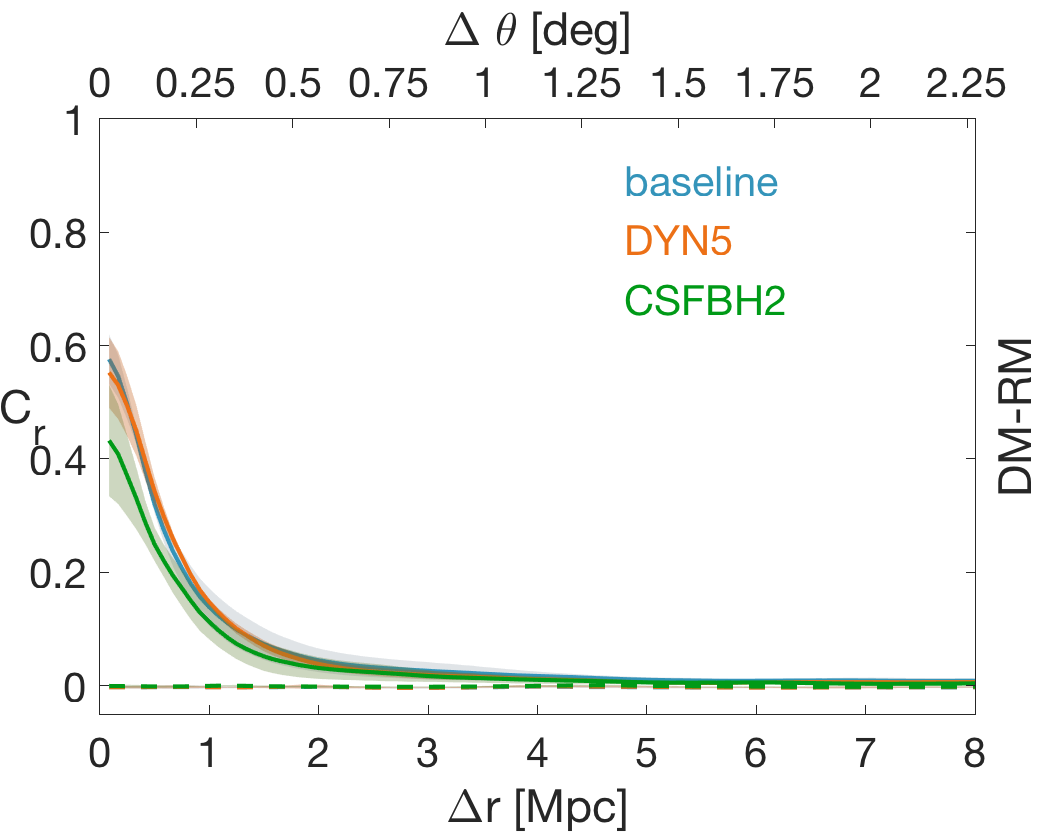}
\includegraphics[width=0.23\textwidth]{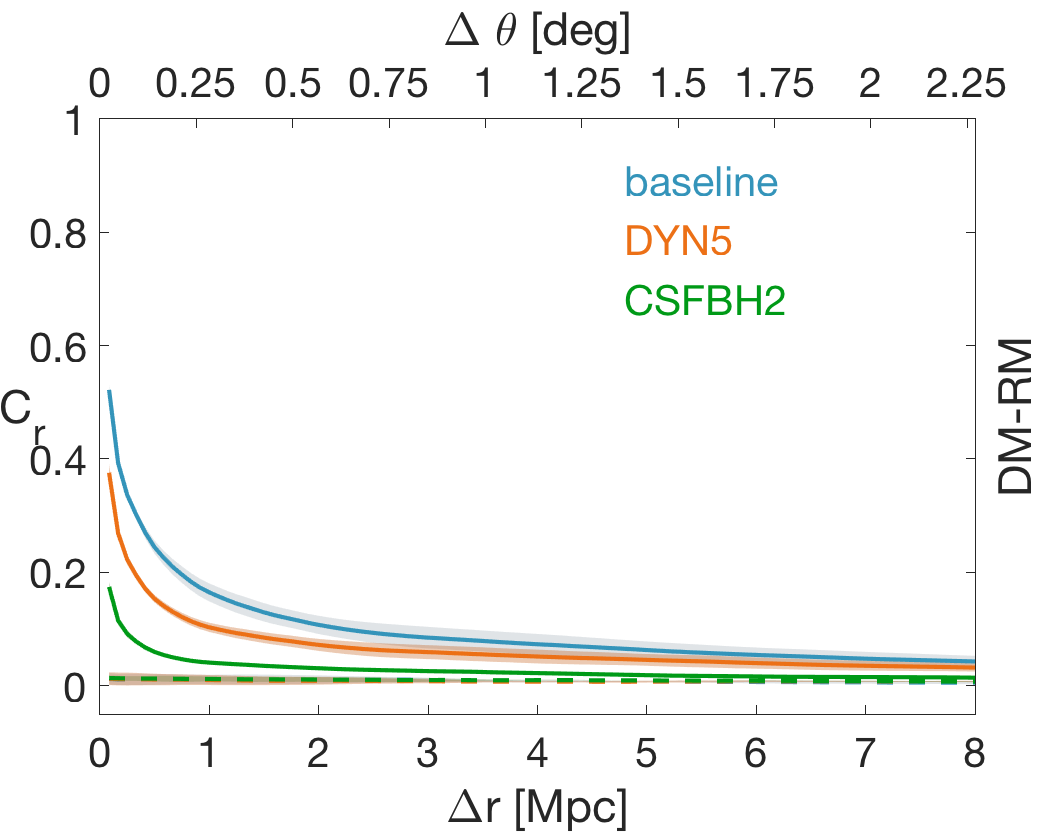}
\includegraphics[width=0.23\textwidth]{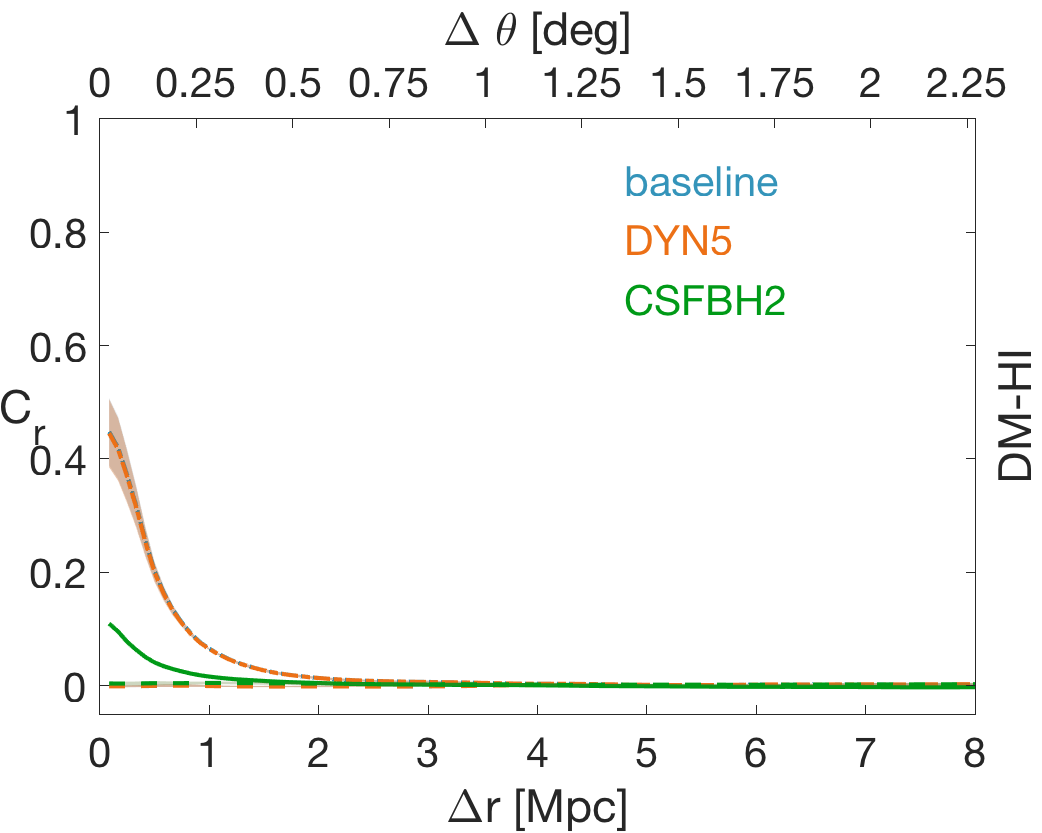}
\includegraphics[width=0.23\textwidth]{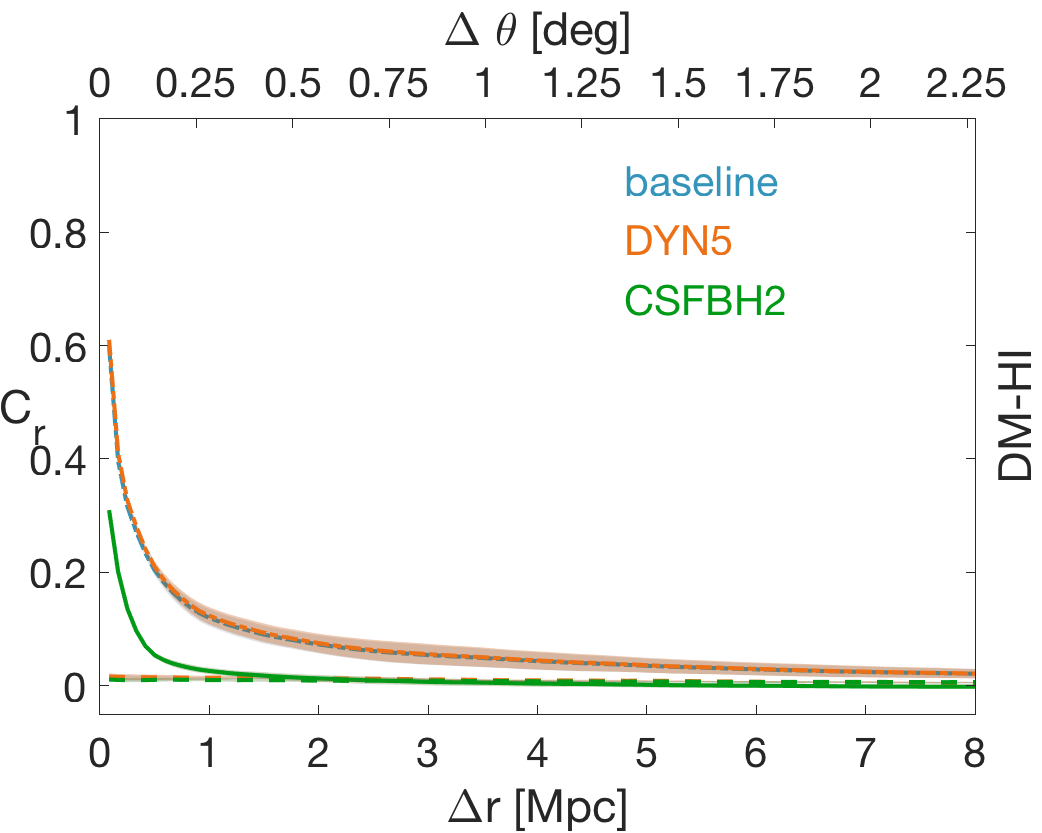}
\includegraphics[width=0.23\textwidth]{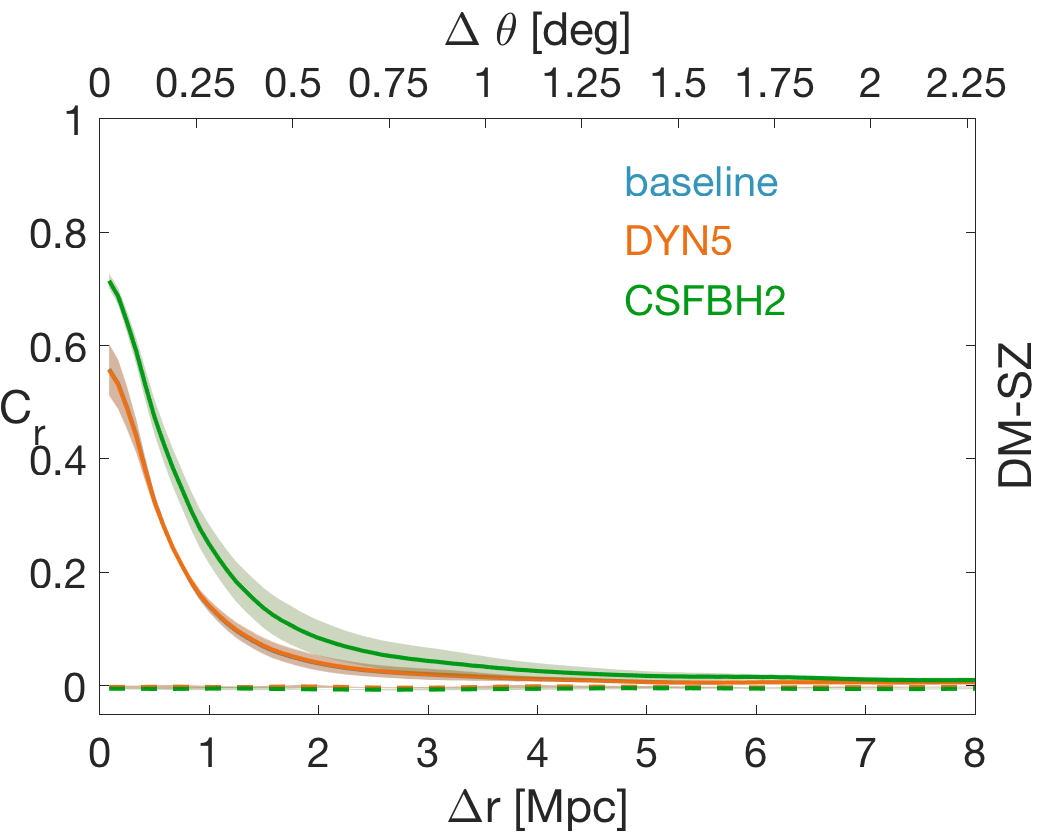}
\includegraphics[width=0.23\textwidth]{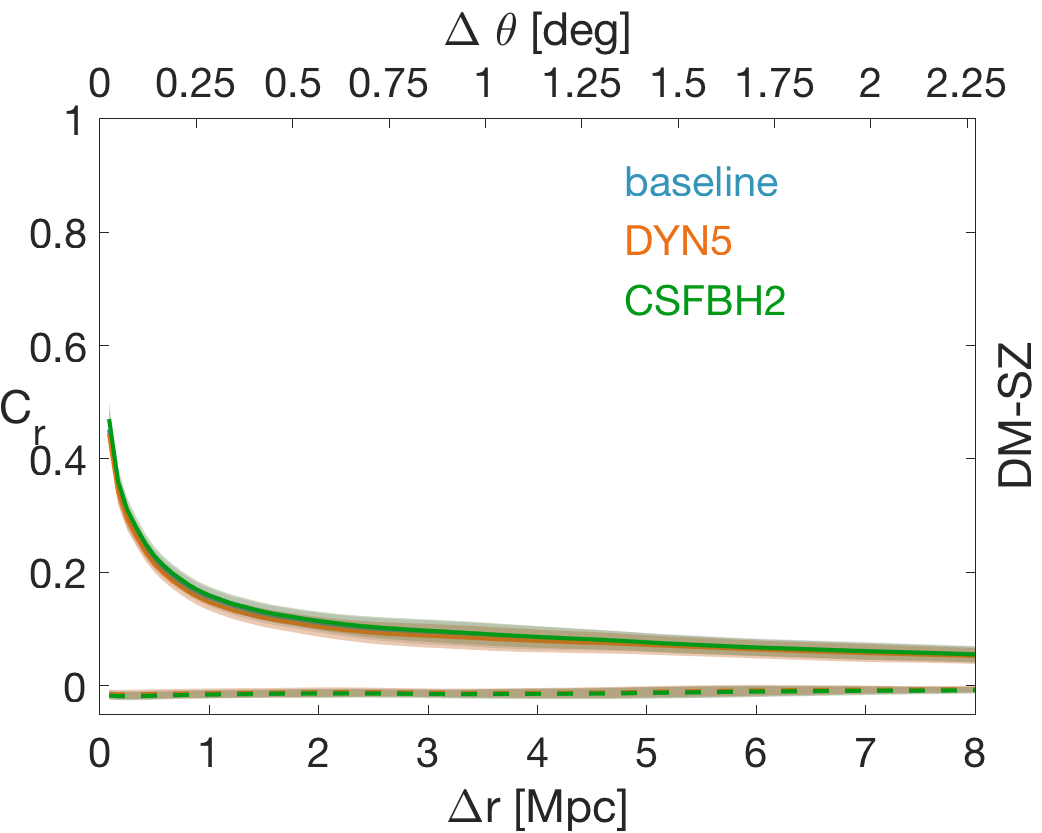}
\includegraphics[width=0.23\textwidth]{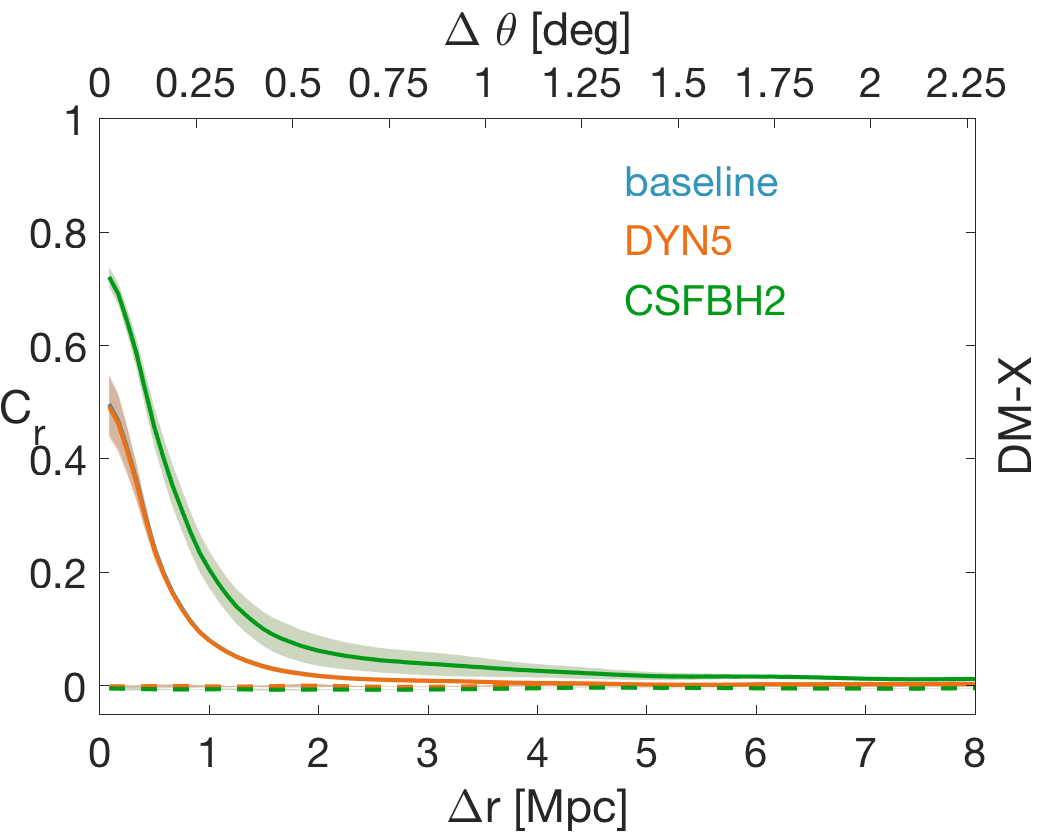}
\includegraphics[width=0.23\textwidth]{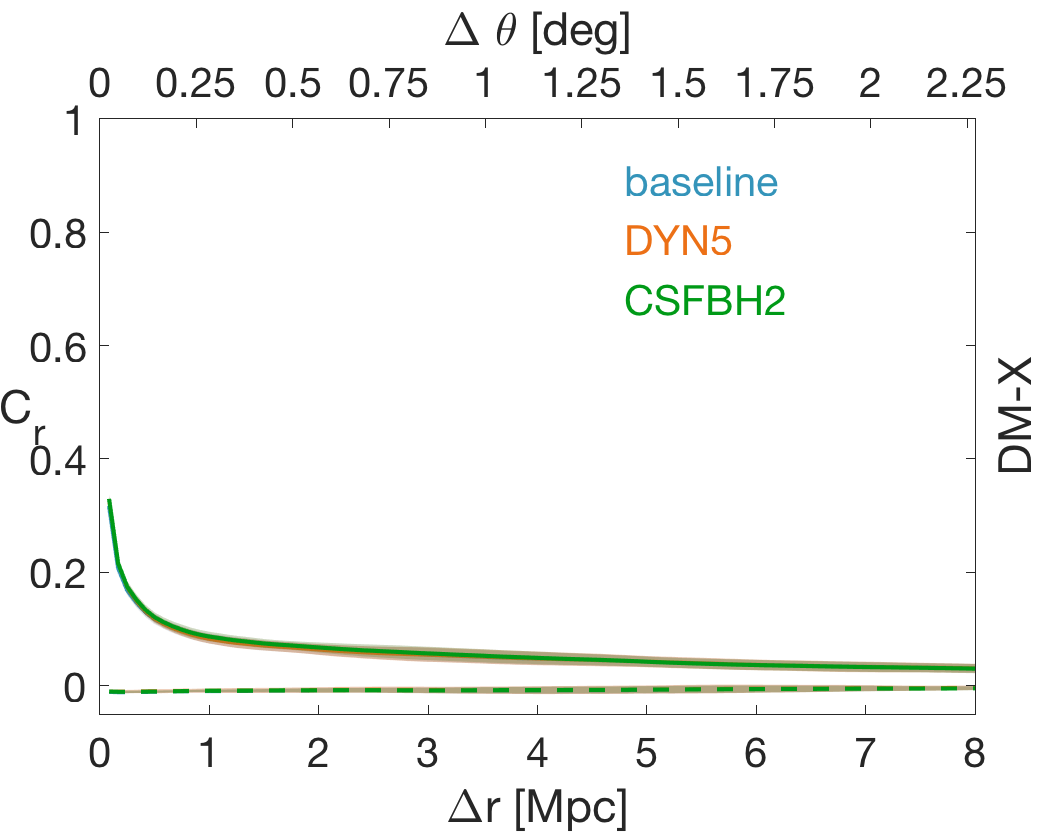}
\caption{Cross-correlation  of the Dark Matter mass distribution and the synchrotron emission (first row), the Rotation Measurement (second row), the HI temperature (third row), the SZ effect (fourth row) and the X-ray luminosity (bottom row). The Cross-correlation is normalised to the corresponding null model. The first column takes into account the unmasked images, while in the second column galaxy clusters and halos have been masked at R100. The coloured shaded bands represent the 1$\sigma$ statistical uncertainty in the calculation of the Cross-correlation, while horizontal dashed lines give the reference correlation of the null model with its 1$\sigma$ statistical uncertainty. In the base horizontal axis, the scale is in Mpc, in the upper horizontal axis the scale is in degrees.}
\label{fig:pure}
\end{figure}

Figure \ref{fig:pure} gives an overview of the the cross-correlation of the five observables, synchrotron emission, RM, HI temperature, SZ effect and X-ray luminosity with the DM mass distribution. The left panel gives the cross-correlation between unmasked maps, while the right panel shows the $R_{\rm 100}$ masking case (all regions containing halos or clusters are masked out from the calculation, see Section \ref{sec:masking}). The correlation curves and their variance, are calculated averaging along the three orthogonal projections, considered as independent. 

All cross-correlations show a significant signal at least out to $\sim 1 ~\rm Mpc$, which corresponds to an angular scale of $\theta \sim 17'$ at $z \approx 0.045$. The cross-correlation between the DM and the synchrotron radiation presents the most peculiar features. The unmasked data show that, differently from all the other correlations, DM and radio emission have a maximum correlation not at displacement 0, where the mass density peaks, but at a distance between 1 and 2 Mpc. There, strong accretion shocks develop, compressing the gas, which increases the intensity of the magnetic field, and accelerating cosmic-rays, leading to an overall enhancement of the radio signal.
When masking is applied, the highest density regions, in which the behaviour of mass and radio emission depart from each other, are removed from the statistics and the cross-correlation tends to follow a monotonically decreasing (with increasing distance) trend. Although the signals are well above the null model, their absolute values are one order of magnitude lower than those found for the other quantities. This is understood because, unlike all other observable, the radio emission is in our model is not a continuous function of the gas density field, but gets ``lighted on" only in shocked cells, hence not all pixels in our sky model contain synchrotron emission, introducing gaps in our radio maps. This also leads to the estimated large variance, which can be interpreted as a projection effect, strong radio emissions spots randomly falling on the same line of sight as the mass density peak. The variance in fact is much lower for the masked data, in which the highest density peaks have been removed.

It is interesting to notice that, in the case of unmasked data, the strongest correlations are found for the DYN5 and the CSFBH2 models, while the baseline model show the weakest signal. By construction, the average magnetic field  strength in the central regions of our halos is similar in all models, but runs with  dynamo amplification or injection by AGN produce a larger spread of the magnetic field magnitude, which is amplified in the synchrotron signal (which approximately scales as $\propto B^2$). Therefore, in these two models the cross-correlation between radio emission and the dark matter distribution is slightly boosted at small separation, considering that most of the signal is originated by halos and by the merger shocks their contain.  The trend gets similar at large distances in all models, albeit with slightly higher signal in the CSFBH2 model, due to the presence of extra magnetic fields and shock waves from satellite halos (and the AGN they contain).  The opposite trend is found for the masked results, i.e. in the baseline model the cross-correlation signal is higher, because in this case similar large scale accretion shocks are present in all models, and the average magnetic field strength remains higher in the baseline one (due to the much higher primordial value), while it sharply drops in the other two scenarios.

For similar reasons, a significant cross-correlation between the RM and the projected DM distribution is present out to a scale of $\sim 2 \rm ~Mpc$ in the baseline model, even with the $R_{\rm 100}$ masking (and likewise for the radio emission). Conversely, the correlation drops below significance at approximately half of this scale for the DYN5 model, and even at shorter distances for the CSFBH2 model.

All the cross-correlations of the remaining quantities with the DM decrease with increasing displacement. The cross-correlations involving the hot gas distribution (X-ray and SZ) are rather similar. The baseline and DYN5 models have lower correlations compared to the astrophysical model, but this difference tends to disappear in masked maps.
Even when halos are masked out,  the cross-correlation signal is significant at the $\geq 1 \sigma$ level out to $\sim 1 \rm ~Mpc$ both in X-ray and SZ. On the other hand, in the unmasked analysis we systematically measure a higher signal for $\leq 2-3 \rm ~Mpc$ in the CSFBH2 model, which is understood because in radiative simulations halos tend to have higher central concentration and correlate slightly more with the DM distribution \citep[e.g.][]{teyssier11}. This is in line with the simulated cross-correlation analysis between thermal SZ and gravitational lensing \citep[][]{2015ApJ...812..154B}, in which the impact of AGN feedback is limited to $\theta \leq 10'-20'$. 

Significant differences between models are found by cross-correlating the HI temperature and the projected DM density distribution, with the non-radiative runs showing the largest correlation at all scales. However, as we noticed in Section \ref{sec:images}, in such runs our assumed fixed ($10^{-6}$)  neutral hydrogen fraction clearly gives a gross overestimate of the HI abundance in the hottest gas phase of the cosmic web, hence only the CSFBH2 models is to be considered realistic here, and in this case a significant excess in the cross-correlation between the projected DM density and the HI temperature decrement is significant only up to $\sim 1 \rm ~Mpc$, also with the masking of our sky model up to $R_{\rm 100}$.\\

In summary, at least in principle, the statistical correlation between the halo/DM distribution and radio observables is a promising tool to probe the amplitude of extragalactic magnetic fields, even outside of the cluster volume usually explored by existing radio observations. This will offer a powerful tool to tell competing models of magnetogenesis apart. On the other hand, observables related to thermal gas are well correlated with the DM/galaxy distribution out to several $\rm Mpc$, with little dependence on the underlying AGN activity. 

\subsubsection{Cross-correlation between other observables}

\begin{figure}
\includegraphics[width=0.24\textwidth]{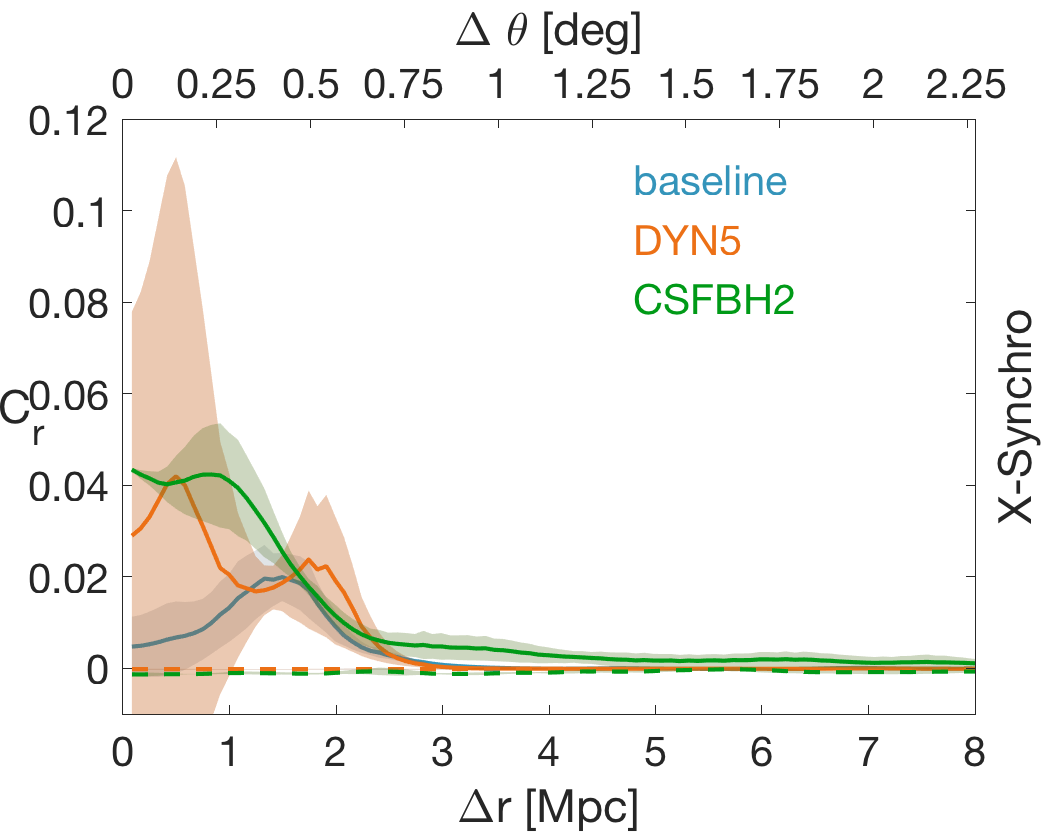}
\includegraphics[width=0.24\textwidth]{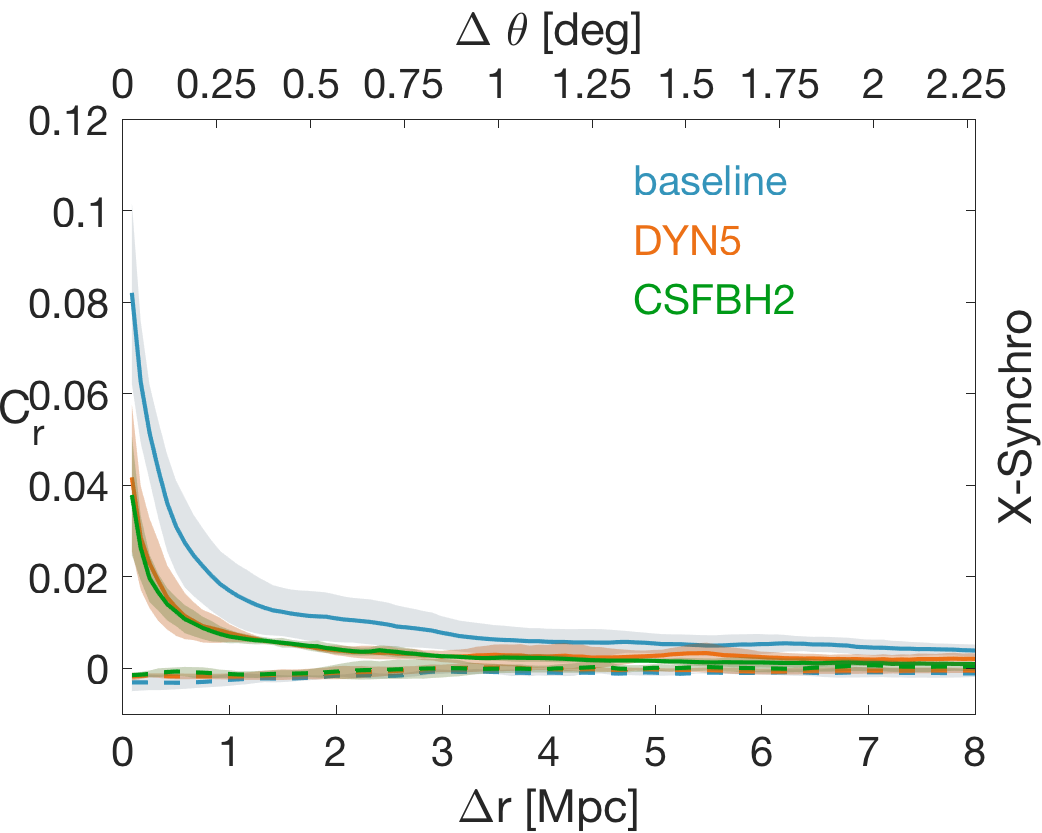}
\includegraphics[width=0.24\textwidth]{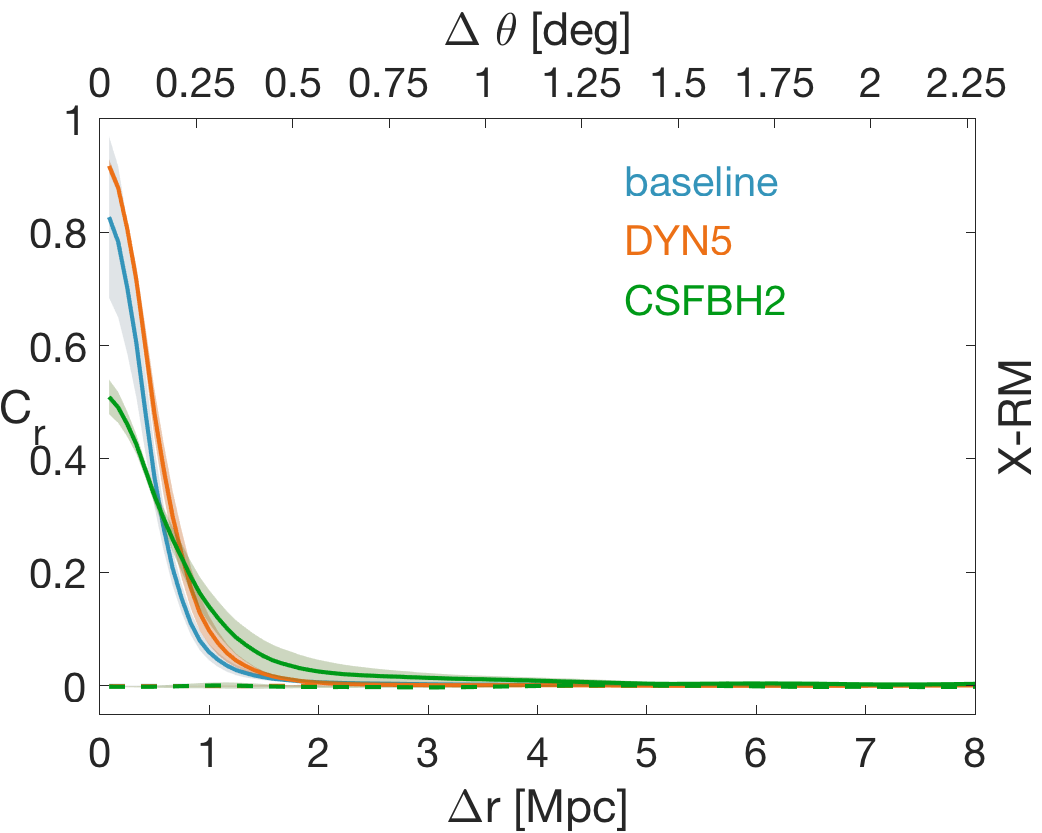}
\includegraphics[width=0.24\textwidth]{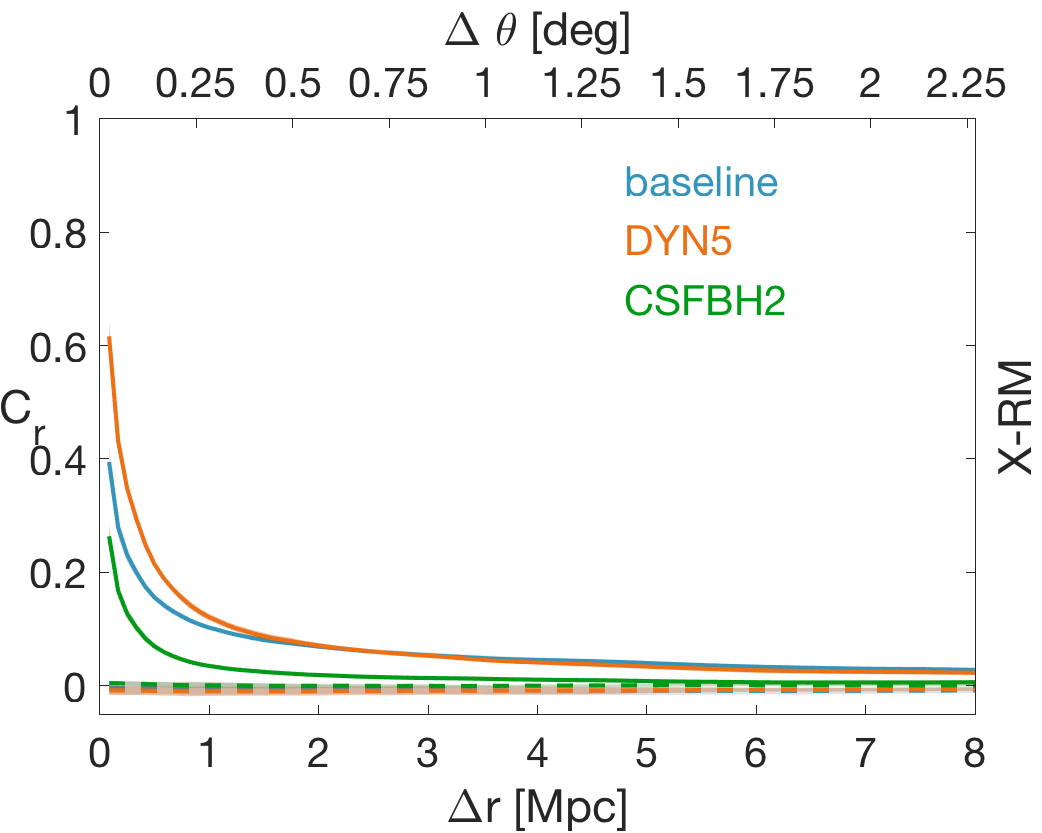}
\includegraphics[width=0.24\textwidth]{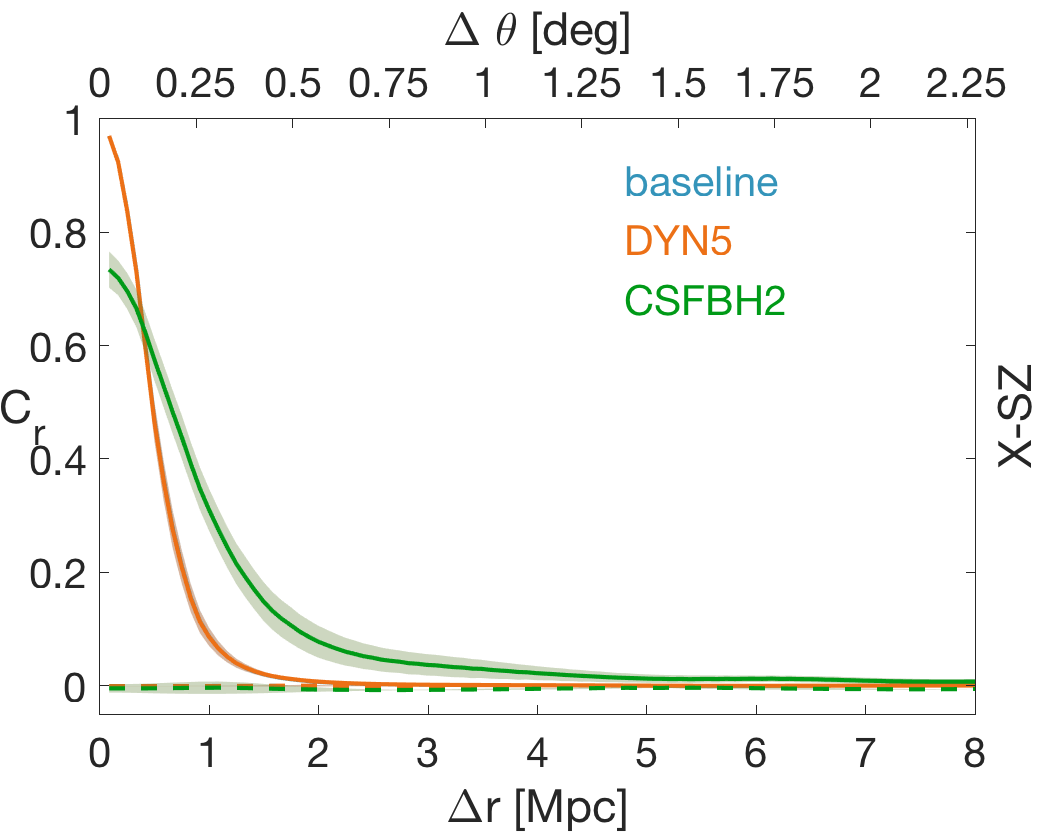}
\includegraphics[width=0.24\textwidth]{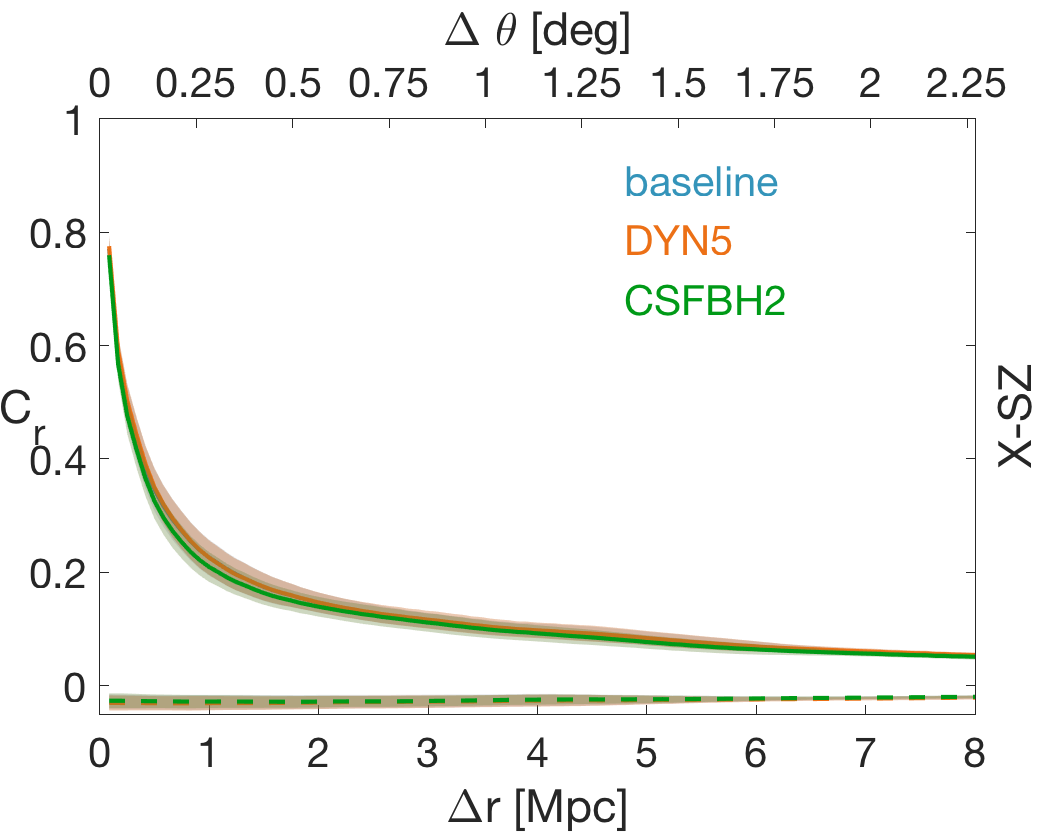}
\includegraphics[width=0.24\textwidth]{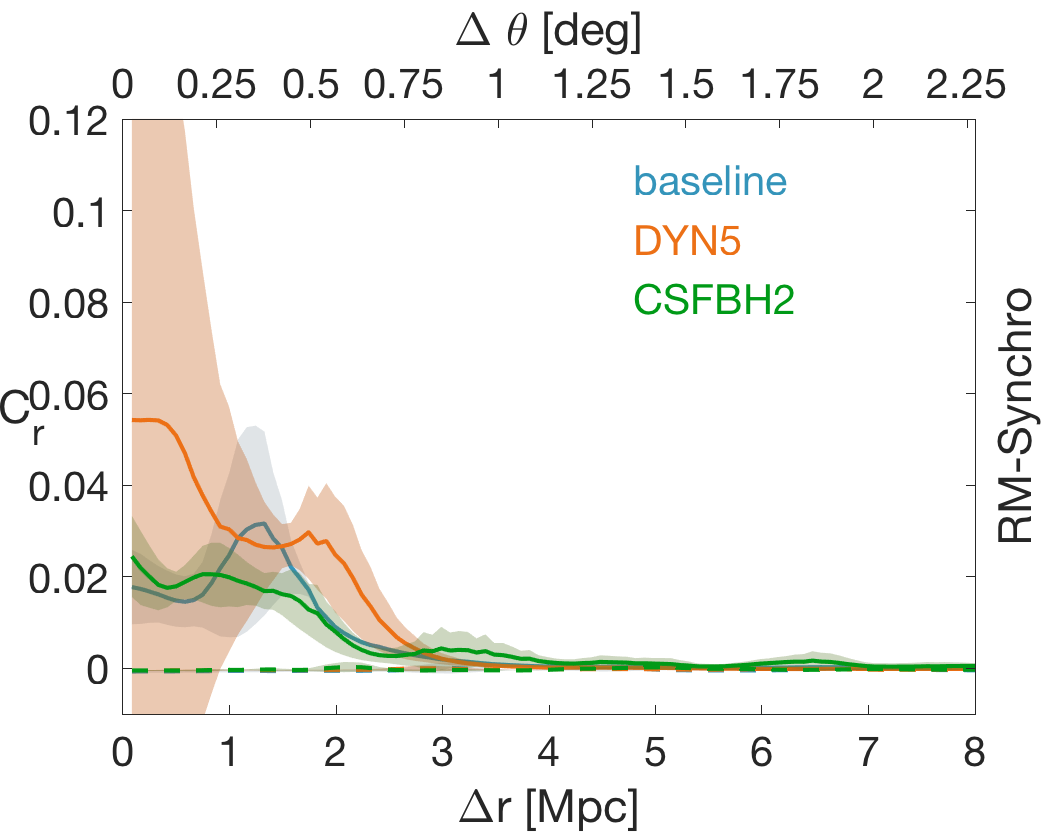}
\includegraphics[width=0.24\textwidth]{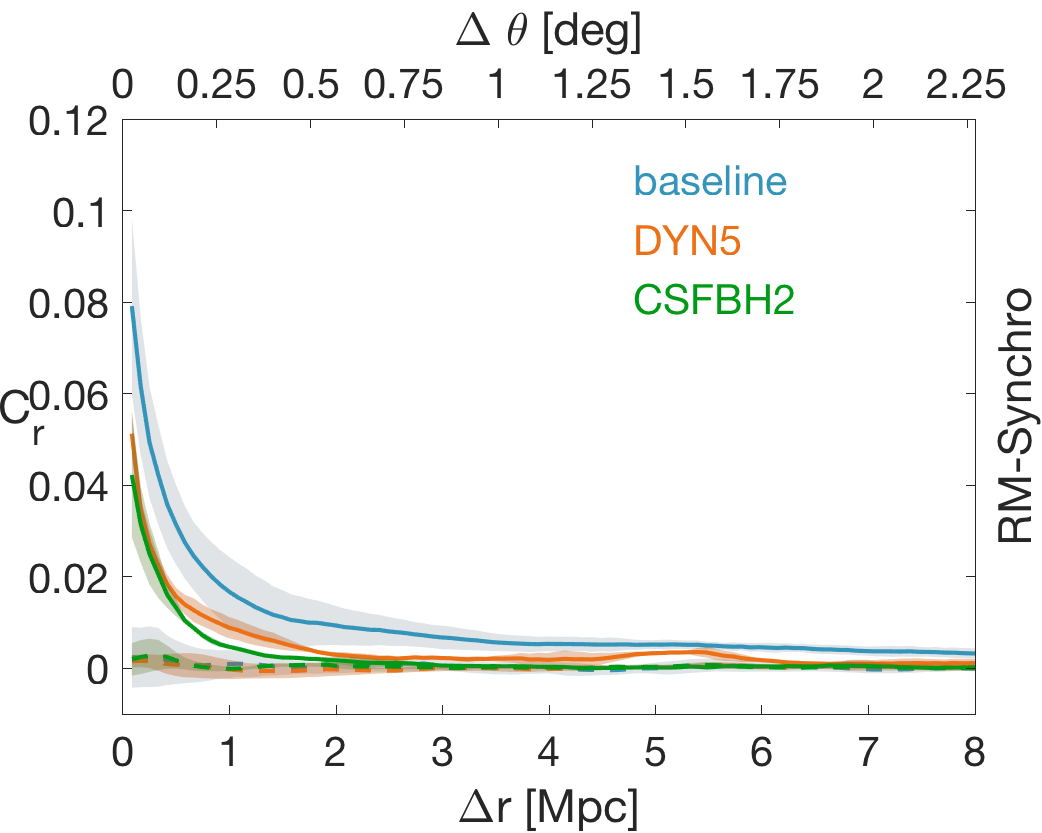}
\includegraphics[width=0.24\textwidth]{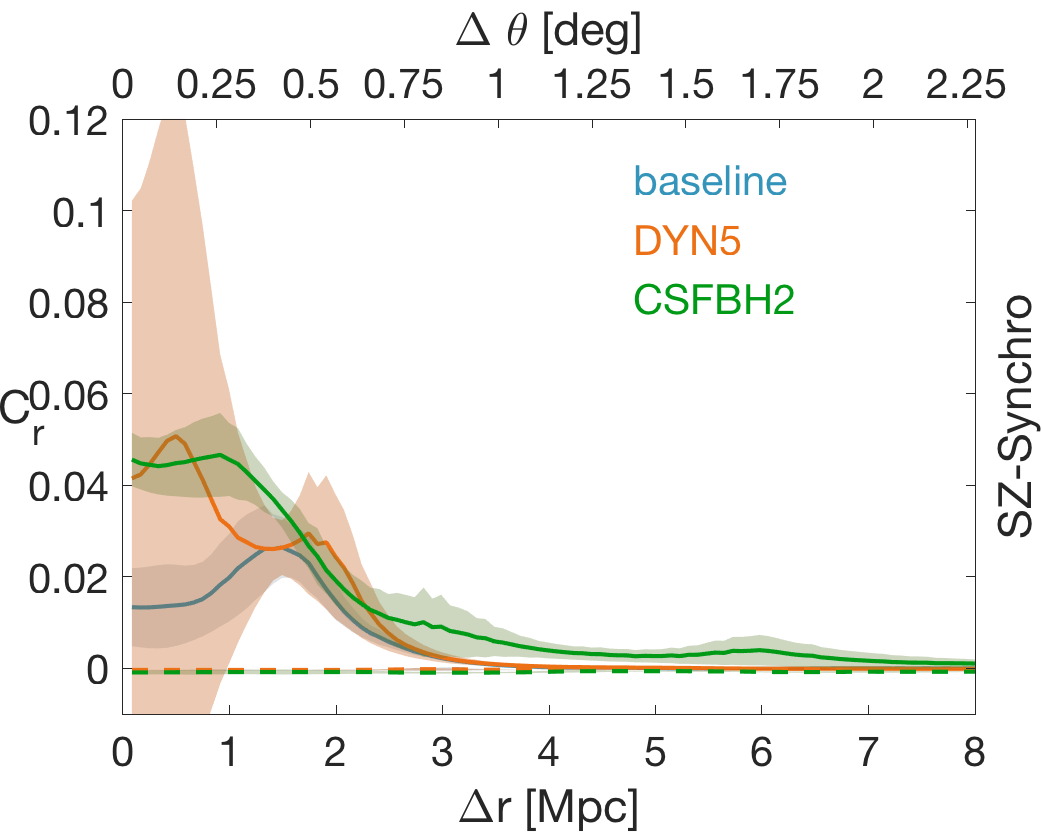}
\includegraphics[width=0.24\textwidth]{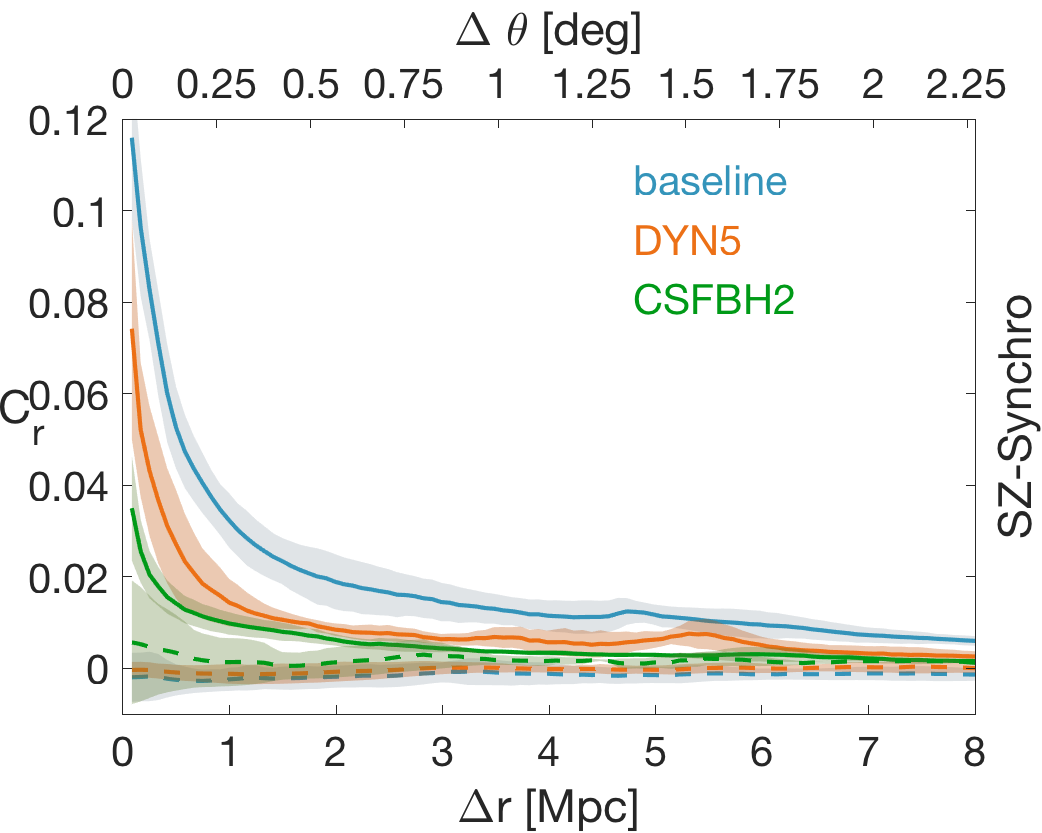}
\caption{Cross-correlations between gas-related quantities. The first row accounts for the unmasked images, while in the second column galaxy clusters and halos have been masked at R100. The coloured shaded bands represent the 1$\sigma$ statistical uncertainty in the calculation of the Cross-correlation, while horizontal dashed lines give the reference correlation of the null model with its 1$\sigma$ statistical uncertainty. In the base horizontal axis, the scale is in Mpc, in the upper horizontal axis the scale is in degrees.}
\label{fig:pure_other1}
\end{figure}

In Figure \ref{fig:pure_other1} we present a selection of the cross-correlations between different gas-related quantities.  While they do not show outstanding differences compared to the trends outlined above, we report them for completeness. Furthermore, they provide ``first order" guidelines for possible future attempts of adopting the cross-correlation analysis for instruments different from those considered in this work. 

With the exception of synchrotron emission, all other gas related quantities strictly follow the total mass distribution and cross-correlate at small displacements.

The cross-correlations between X-ray and RM or SZ and between RM and SZ have the highest and cleanest signal, hence they appear to be the most promising for detection. On the other hand, the cross-correlations between synchrotron radio emission and SZ, X-ray emission and RM tend to be weaker in absolute value (owing to the intermittent nature of radio emission in our model, as above) but spatially broader, with the extreme case of baseline model in which a significant cross-correlation with all quantities can be measured out to  $4-5$ Mpc. In the case of the radio sky model, large statistical errors affect the unmasked data (following the rare occurrence of central merger shocks), while masked data results to be less noisy. The signal is low (always below 0.1), but significantly higher than the null model, especially in the correlations with X and RM.\\

Finally, all cross-correlations involving HI in the CSFBH2 model (not shown), the only one treating self-consistently the HI component, are very weak, with trends similar to the cross-correlation with DM presented in the previous section. Little correlation is found even at null displacement, which is expected because in the CSFBH2 model (the only one in which a basic chemical evolution model is adopted at run time) HI almost never forms within hot and massive halos, hence a large SZ and X-ray signal anti-correlates with HI temperature. For unmasked data, the cross correlation tends to zero at scales between 1 and 2 Mpc, being dominated by the collapsed, high-density structures. Such correlation length tends to be larger for masked data, since random correlations are statistically more frequent considering the filamentary large scale distribution of matter only.


\begin{figure*}
\includegraphics[width=0.9\textwidth]{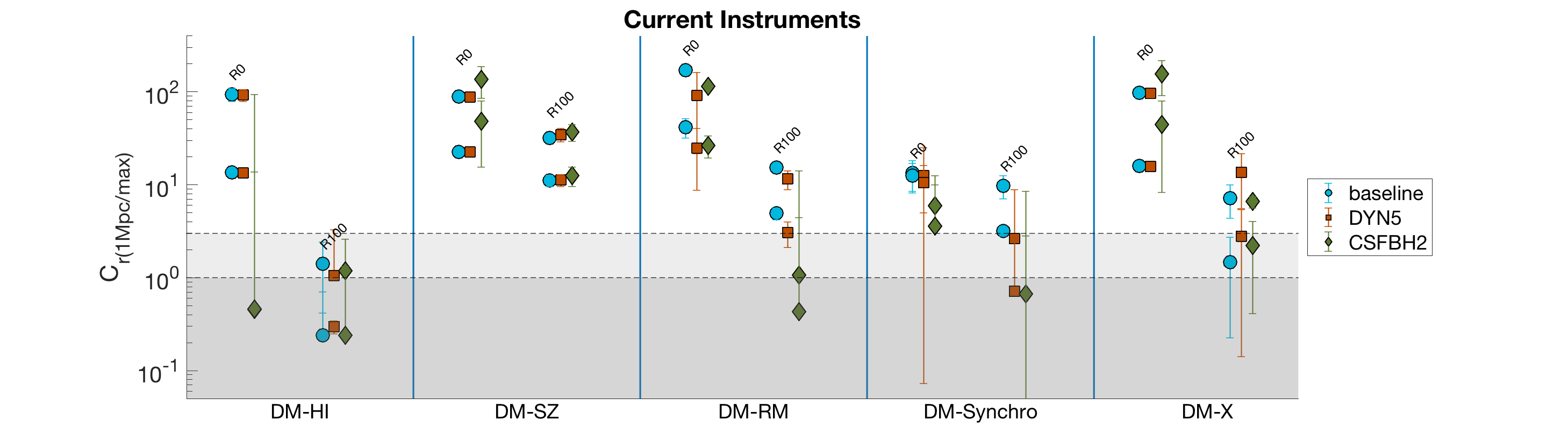}
\includegraphics[width=0.9\textwidth]{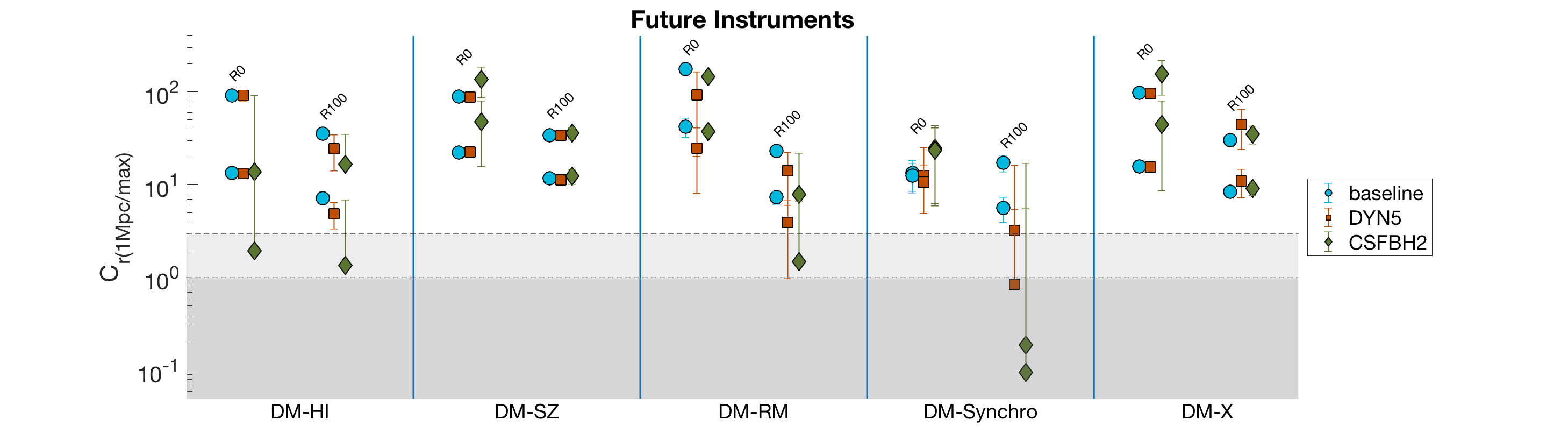}
\caption{Cross-correlation of Dark Matter with HI temperature (first block - each block is the area delimited by vertical blue lines), SZ (second block), Rotation Measurement (third block), synchrotron emission (fourth block) and X-ray luminosity (fifth block), in presence of noise due to different instruments. The top panel shows the results for ``current'' instruments and the bottom panel for ``future'' instruments (see Table \ref{table:noises} for details). Within each block, each group of three represents a different masking. Within each group of three, left is the baseline model (blue circles), centre is the DYN5 model (orange squares) and right is the CSFBH2 model (green diamonds). Top symbol is the maximum correlation, bottom symbol is the correlation at 1 Mpc displacement ($\approx$ 17 arcmin separation). Horizontal grey stripes show the null model 1$\sigma$ and 3$\sigma$ uncertainty. Error bars represents the standard deviation of the cross-correlation function calculated over the three orthogonal projections.}
\label{fig:dm-noise}
\end{figure*} 

\begin{figure*}
\includegraphics[width=1.1\textwidth]{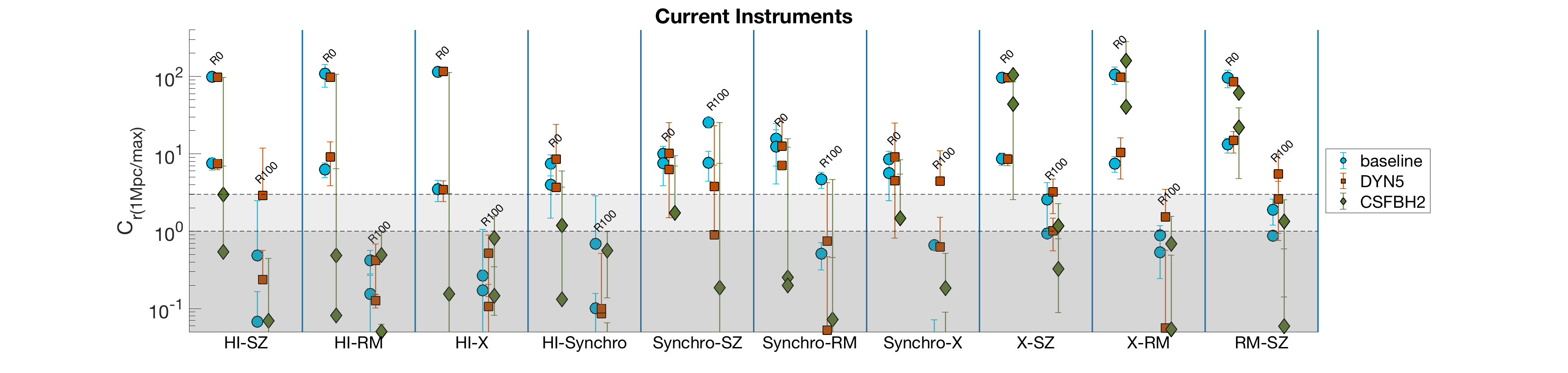}
\includegraphics[width=1.1\textwidth]{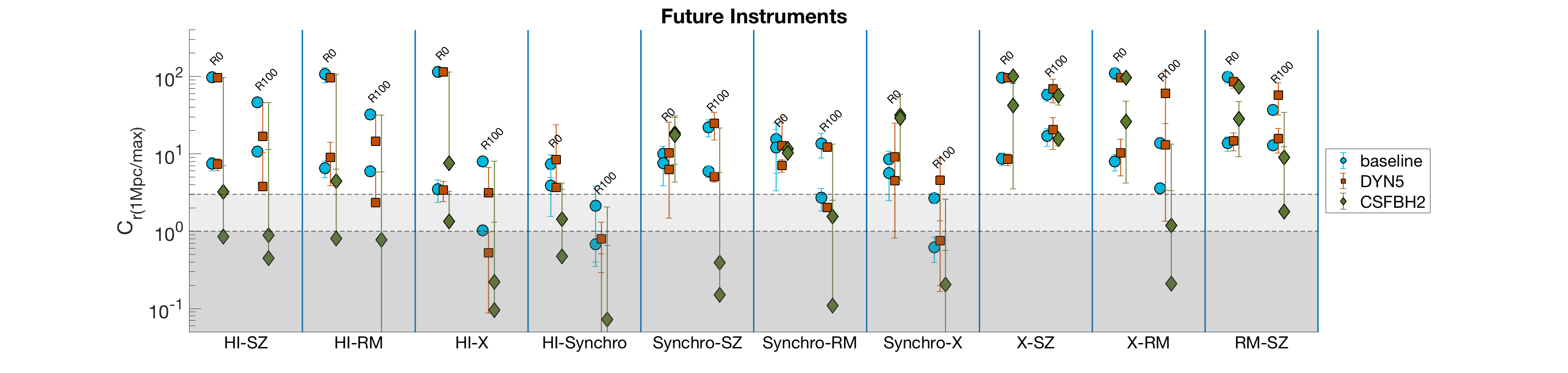}
\caption{Cross-correlation of various gas related quantities among each other. HI temperature with SZ (first block - each block is the area delimited by vertical blue lines), RM (second block), X-ray luminosity (third block) and synchrotron emission (fourth block), synchrotron emission with SZ (fifth block), RM (sixth block) and X-ray luminosity (seventh block), X-ray luminosity with SZ (eighth block) and RM (ninth block), RM and SZ (last block) in presence of noise due to different instruments. The top panel shows the results for ``current'' instruments and the bottom panel for ``future'' instruments (see Table \ref{table:noises} for details). Symbols and grouping follow the same rules as in Figure \ref{fig:dm-noise}.}
\label{fig:others-noise}
\end{figure*} 

\begin{table*}
\begin{center}
\caption{List of observable properties and adopted detection threshold. We associate to each observable a reference instrument, which loosly corresponds to the adopted observational cut (see ``note" in the table for more information).}
\footnotesize
\centering \tabcolsep 2pt
\begin{tabular}{c|c|c|c|c}
  observable & frequency/en.range & instrument & detection threshold & note  \\ \hline
  X-ray emission   & 0.3-2.0 keV & eROSITA & $\approx 2 \cdot 10^{-15}  \rm erg/s ~cm^2$ & 10 ks (polar region survey)\\ 
X-ray emission  & 0.3-2.0 keV & ATHENA-WFI & $\approx5 \cdot 10^{-16}  \rm erg/s ~cm^2$ & 100 ks  \\  \hline
    differential HI Temperature & 1400 MHz & ASKAP & $\approx10^{-3} \rm ~K$ & Possum\\ 
    differential HI Temperature  & 1000 MHz & SKA-MID & $\approx10^{-5} \rm ~K$ & Phase II survcey \\ \hline
    radio emission & 200 MHz & LOFAR-HBA & $\approx1.0 \rm \mu Jy/arcsec^2$ &  Tier I survey\\ 
    radio emission & 200 MHz & SKA-LOW & $\approx0.2 \rm \mu Jy/arcsec^2$ & 2 years survey\\  
     radio emission & 180 MHz & MWA  & $\approx 0.28 \rm \mu Jy/arcsec^2$ & MWA Phase I\\  \hline
    Faraday Rotation & 1400 MHz & ASKAP/Meerkat & $\approx1~ \rm rad/m^2$  & Possum/Mightee-Pol \\ 
    Faraday Rotation & 1000 MHz & SKA-MID& $\approx0.1~ \rm rad/m^2$ & Phase II\\  \hline
    Compton Y-param. & 220 GHz & PLANCK & $\approx 2890$  & \\
    Compton Y-param.  & 21 GHz & AtlAS & $\approx 1000 $  & \\ \hline 
     
  \end{tabular}
  \end{center}
\label{table:noises}
\end{table*}    

\subsection{Cross-correlation analysis including observational effects}

In this section, we focus on the most significant selection of cross-correlations between our observables, with observational/instrumental noise added. The purpose is that of assessing which signatures of the cosmic web may be potentially detectable with present-time surveys, in the near or in the more remote future.

In detail, we first consider a representative set of currently available instruments at all wavelengths (e.g. eRosita\footnote{https://www.mpe.mpg.de/eROSITA}, ASKAP\footnote{https://www.atnf.csiro.au/projects/askap/news.html}, LOFAR-HBA\footnote{https://www.astron.nl/telescopes/lofar}, Meerkat\footnote{https://www.sarao.ac.za/gallery/meerkat/} and PLANCK\footnote{https://sci.esa.int/web/planck}), while for future instruments we assume the performance of a few notable planned/proposed instruments (e.g.  ATHENA-WFI\footnote{http://www.mpe.mpg.de/ATHENA-WFI/}, SKA-MID and  SKA-LOW\footnote{https://www.skatelescope.org} and AtLAST \footnote{http://atlast.pbworks.com/}). 
Our approach here is to refer to the nominal sensitivity/performance of the various surveys, as presented in their reference papers and/or official websites. In most cases, these sensitivities can be reached in the search of diffuse emission only in the ideal case of a full removal of pointlike sources and foreground emissions as well as under the assumption of a perfect calibration of the instruments. 

For the projected galaxy distributions, we set the threshold of projected DM density so that it is compatible with the present sensitivity of the WISE IR survey, yielding $\sim 10$ galaxies per square degree for $z \leq 0.07$  \citep[][]{vern17}, and the future sensitivity by EUCLID, which should give $\sim 10^3-10^4$ galaxies per square degree in the same redshift range \citep[e.g.][]{2012MNRAS.427.3134B}. 

We should preliminary notice that producing synthetic observations convolved for the specific resolution of each different instrument is out of the scope of the paper. 
Here we mostly target emission or absorption features from large scales of the cosmic web.
Our pixel resolution  ($90\rm "$) is thus coarser than what can be achieved by several instruments. 
One example is the Rotation Measure of background polarised radio sources, which can get to $\sim 10"$.  However, we do not consider this problematic for our science goal here, as recent papers have shown that the simulated RM only drops by a factor of a few when large beams are applied to convolve the simulated radio signal \citep[][]{va18mhd,wi19}.  Moreover, previous resolution studies have shown that the 3-dimensional properties of the magnetic field in filaments and/or outside galaxy clusters are overall little affected by resolution \citep[e.g.][]{va14mhd}. Therefore our predictions here are reasonably accurate (within a factor of $\sim 2-3$ on RM) in the case of the diffuse intergalactic medium of the cosmic web. 

As an exception, in the case of PLANCK SZ observations the resolution beam is $\approx 5'$, coarser then our mock sky model. However, our simulations do not show much structure in the SZ from the cosmic web on large scale, hence the cross-correlation analysis we present here is reasonably robust against modest inhomogeneities in the simulated versus real beam size of observations. 

Only for the specific comparison with MWA observations, discussed in Section \ref{sec:mwa}, we tuned the spatial resolution of the mock observation in order to compare more closely with recent observations.\\

The two panels of Figure \ref{fig:dm-noise} give a synthetic overview of the predicted cross-correlation signal for current and future instruments. We give the values of the amplitude of the cross-correlation at null displacement or at the fixed reference displacement of 1 Mpc, both for the unmasked and for the R100 masked models. For a homogeneous presentation of data, we normalised the cross-correlation signal at these two reference separations by the maximum cross-correlation of the corresponding null model, which now accounts for the contribution of both the cosmic variance and the noise assumed in each specific observation. 

Several interesting trends can be noticed:
\begin{itemize}
    \item {\it DM-HI correlation}: the cross-correlation between IGM and dark matter in the most realstic model (CSFBH2) becomes nearly impossible to be detected by present instruments (e.g. Meerkat/ASKAP), while it should be detectable up to $R_{\rm 100}$ using the SKA-MID (mostly in Phase II), in agreement with \citet[][]{2017PASJ...69...73H}. The amplitude of the signal falls rapidly when the clumpiest portion of the sky model is excised. However, the impact of our limited spatial resolution in modelling the formation of HI even outside of halos is yet to be assessed with higher resolution simulations. 
    
    \item {\it DM-SZ correlation}: the cross-correlation holds up to $\sim 1$ $\rm Mpc $ even when halos are masked out, with little dependence on the assumed physical model, with both current and future instruments.
    
    \item {\it DM-RM correlation}: for a $\sim \rm rad/m^2$ sensitivity level, the significant detection of cross-correlation seems possible even when halos are masked out, for a significant scale magnetic field as  in our baseline model. The detection at $R_{\rm 100}$ becomes marginal for the sub-grid dynamo model, and  impossible in the CSFBH2 scenario.  However, a ten-fold increase in RM sensitivity, as expected to be possible with the SKA-MID in Phase II, may allow detecting RM outside of halos even in the CSFBH2 model, and hence discriminate between magnetogenesis scenario using RM grids;
    
    \item{\it DM-Synchrotron correlation}: similar to the previous case, but with somewhat lower significance, detections are possible outside of halos in the primordial case, with a sensitivity of order $\sim \rm \mu Jy/arcsec^2$ as in LOFAR-HBA. Detection will be even more clear with the SKA-LOW in this scenario. Also in the dynamo amplification model future SKA-LOW observations should allow to marginally detected a positive cross-correlation with the underlying galaxy distribution. Detecting the signature of radio emission outside of halos in the CSFBH2 model will remain challenging even with SKA-LOW, due to the rapid drop ($P_{\rm radio}\propto B^2$ for $B\ll 3.2 \rm ~\mu G$) of the radio emission away from halos \citep[][]{va17cqg}, in the case magnetic fields are only seeded by processes linked to galaxy formation; 
    
    \item{\it DM-X-emission correlation:} when the correlation at 1 Mpc is concerned the robust detection of the correlated signal from X-ray emission in the soft band appears fully feasible only with future instruments, i.e. with $\sim 100$ ks integration with Athena-WFI (in line with \citealt[][]{vazza19}). Statistical detections using present instruments, like eRosita, are extremely challenging, with little dependence on the assumed gas physics. Thus for a proper imaging of the WHIM in the cosmic web, a new concept of X-ray telescope must be deployed, for which proposals have been submitted \citep[e.g.][]{2018arXiv180909642T,2019arXiv190801778S}. 
    
\end{itemize}

In summary, with presently available surveys of galaxies and with current multi-wavelength instruments, the best chances of detecting the correlated signal of the diffuse IGM outside of halos and in filaments comes from SZ observations (regardless of adopted gas physics, e.g. \citealt{2010MNRAS.401.1670F,2015ApJ...812..154B}). 

Additional chances of detecting the magnetised cosmic web in correlation with the galaxy distribution may come from surveys of Rotation Measure, in case of the significant volume-filling magnetic fields ($\geq 1-10 \rm ~nG$) expected from a primordial scenario as in our baseline model. Conversely, a robust non detection of such correlated signal with surveys of RMs can already restrict the allowed amplitude of primordial magnetic fields, at the $\leq \rm ~nG$ level. 

However, in practice the effective sensitivity of any RM survey can be limited due the contribution from the foreground Faraday screen by our galaxy as well as by the intrinsic RM from polarised background sources, both challenging to remove \citep[see discussion in][]{lo18}. By studying the dependence with redshift of RM from a quasar sample,  \citet{2017ARA&A..55..111H} concluded that $\sim 10^4-10^5$ measured RMs may be necessary to tell apart Galactic from extragalactic contribution in such objects. The ever growing knowledge of the three-dimensional structure of the Galactic magnetic field should also improve alongside the growth of RM samples, enabling the removal of the Galactic foreground, in combination with other observables (such as extragalactic RMs, PLANCK polarisation data, galactic synchrotron emission and observed distribution of ultra-high energy cosmic rays, see 
\citet[][]{2018JCAP...08..049B} for a recent review).
Therefore, the theoretical RM sensitivity that should be reached by the SKA-MID ($\sim 0.1$ $\rm rad/m^2$) is a very optimistic one, which can only be achieved in presence of major advances in the modelling of the polarisation sky. 

On the other hand, the somewhat reduced significance of the correlation between DM and synchrotron emission in most models should be balanced by the fact that it is comparatively easier to remove the foreground contribution to the radio sky (e.g. based on the spectral index of the observed emission), and that the emission from radio galaxies is generally well confined in host clusters/groups, and is hence enclosed within the masked areas. 
Therefore, the challenging statistical detection of the cosmic web in total radio intensity may offer a strong case for the study of cosmic magnetism. \\

Finally, we considered mixed cross-correlation between observables that directly trace the gas component and/or the magnetic fields, with the same realistic sensitivities considered above, as shown in Fig.\ref{fig:others-noise}. The most promising cross-correlation appears to be between the SZ effect and the synchrotron emission, at least in the primordial scenario. Marginally detectable cross-correlations are present between the SZ effect and RM, especially for the primordial and  for the small-scale dynamo amplification case.

In more futuristic scenarios many of such correlation may become detectable, even at the distance of 1 Mpc and adopting masking.
The  correlation between SZ effect  and synchrotron emission should be prominently detectable in the primordial case, and still marginally detectable in the small-scale dynamo amplification scenario, hence offering a way to measure magentogenesis based on the amplitude of detected (or un-detected) cross-correlation. Likewise, also significant cross-correlations between SZ effect and RM should be detectable for these two scenarios, as well as between X-ray emission and SZ effect. In all cases, the detection of the cosmic web through magnetic related effects depends on the high sensitivity that should be achieved thanks to the full deployment of the SKA, both in its LOW and MID parts. Interestingly, also cross-correlations entirely produced in the radio domain should detect the cosmic web, i.e. through the synchrotron emission-RM correlation, which would be prominent both in the primordial and in the dynamo model, while it should remain undetectable even by the SKA in case of a purely astrophysical origin of magnetic fields. 

\subsubsection{Comparison with MWA Phase I cross-correlation}
\label{sec:mwa}

Finally, we attempt a qualitative comparison with the recent 
results by \citet{vern17}, who have cross-correlated the radio emission in MWA Phase I observations and the galaxy distribution from the WISE+2MASS galaxy survey, for a  $22^\circ \times 22^\circ$ field of view. This work reported no statistically significant detection of cross-correlation on $\geq 20'$ scales, while correlated signal on smaller angular scales is likely due to the contamination of unresolved radiogalaxies within the resolution beam of MWA. 

Here we assumed the same sensitivity and resolution beam quoted by \citet{vern17} for MWA Phase I observations at 180 MHz, as well as lowered the number of detected galaxies in order to mimic the WISE and 2MASS statistics.
In detail,  we convolved our radio sky model using a $\theta \approx 2.9'$ resolution beam (e.g. $\sim 2$ times larger than what we used in the rest of the paper) and considered a noise level of $0.96 \rm ~mJy/beam \approx 0.028 \rm ~\mu Jy/arcsec^2$ at $180$ MHz, corresponding to the deepest MWA (Phase I) observations used in \citet{vern17}. To match the galaxy density used in \citet{vern17}, we used a DM density threshold higher than in the rest of the paper: $\rho_{th} = \sim 6 \cdot 10^{-29} \rm ~g/cm^3$. No masking of halos is used in this case.\\

\begin{figure}
\includegraphics[width=0.5\textwidth]{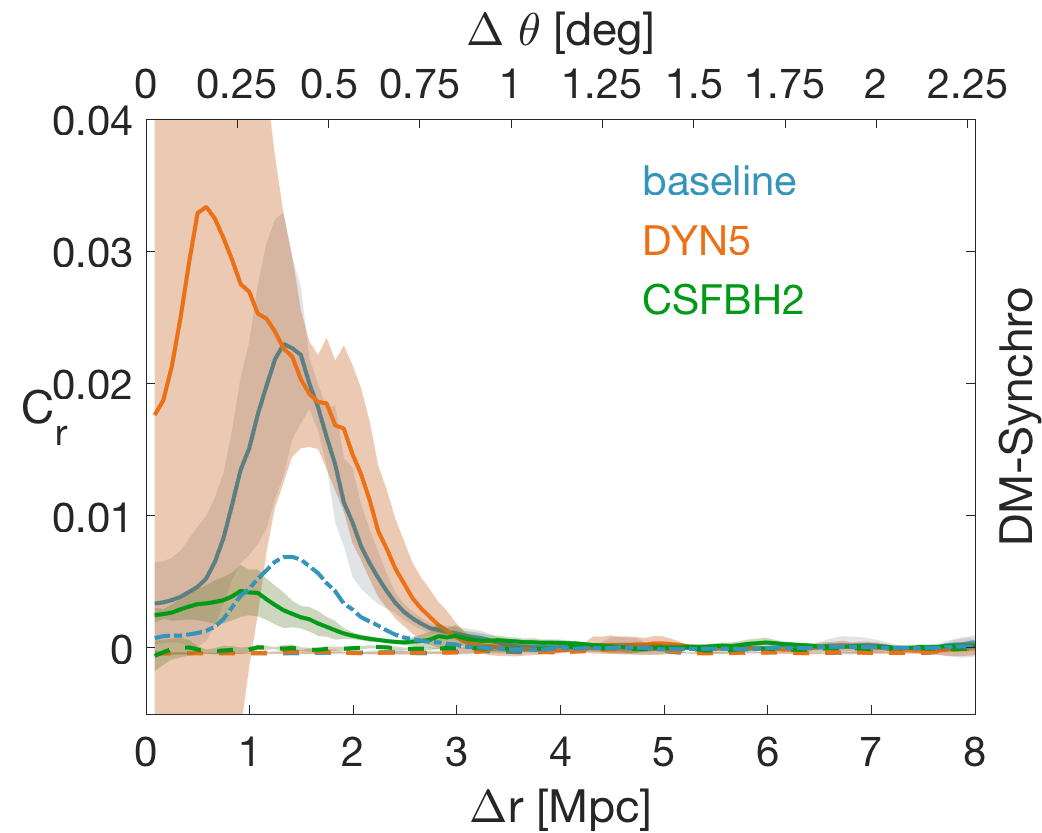}
\caption{Cross-correlation between the DM mass and the synchrotron emission adopting the MWA set-up. Lines and bands have the same meaning as in Figure \ref{fig:pure}. Blue dotted-dashed line represent the baseline model rescaled down 100 times}
\label{fig:mwa}
\end{figure}

Our results are shown in Fig.\ref{fig:mwa}, and shall be compared with \citet{vern17} results for their lowest redshift sub selection of data ($z \leq 0.13$).
We cannot readily compare with the cross-correlation values given by \citet{vern17} in a quantitative way, owing to the different approaches in estimating the noise level of the cross-correlations. Therefore, our synthetic observation can only qualitatively address which model seems to be more compatible with MWA observations.

The dynamo model gives the largest correlation with the galaxy distribution, followed by the baseline model, while the astrophysical scenarios gives the lowest level of cross-correlation. While all models give a significant cross-correlation out to $\sim 40-50'$, the amplitude of correlation in the primordial and in the dynamo models seem to be too large to have been missing missed by MWA observations. This potentially suggests that a $1 \rm ~nG$ initial field  (resulting into a typical magnetisation of filaments of $10-100$ nG as shown in \citealt{gv19}) is too large to be compatible with the lack of detection reported by \citet{vern17}. 
If the radio emission is instead rigidly rescaled by a factor 100 downwards, corresponding to an initial magnetic field in the simulation of $\approx 0.1$ nG comoving, the significance of the
cross-correlation approaches the one of the astrophysical case, showing only a weak correlation excess out to $\sim 20'$.
This test also suggests that the efficient amplification of magnetic fields in our dynamo model,  introduced ad-hoc to mimic the scenario proposed by \citet{ry08} and challenging to directly observe in numerical simulations, is probably not at work in the the bulk of filaments in the cosmic web. 

As a caveat, we must notice that, given the finite mass resolution of our simulations, we cannot properly form dwarf galaxies in voids (or in very poor environment, in general). Therefore, even if the number of galaxies is calibrated to be at the level of the galaxy distribution observed in WISE/MASS surveys, our spatial distribution is typically more clustered than in observations. In principle, this can decrease the cross-correlated signal coming from low density regions in our sample. With future work, we will employ more resolved simulations in order to better address this issue. 
\\

\section{Conclusions}

In this work, we have used recent MHD cosmological simulations to investigate the use of cross-correlation analysis between different observables of the cosmic web (i.e. X-ray emission, Sunyaev-Zeldovich signal at 21 cm, HI temperature decrements, diffuse synchrotron emission and Faraday Rotation).
Our analysis aims at both interpreting already available observational attempts in this direction \citep[e.g.][]{vern17}, and exploring what can be achieved with future multi-wavelength surveys. 

For the sake of performing an homogeneous study of many observables with the same dataset, 
our analysis is bound to oversimplify several aspects, which we shortly discuss here.
The most important limitations of our simulations are related to the effect of the limited spatial resolution on small, high-density peaks in the gas distribution  (i.e. small galaxies or galaxy cluster cores). 

First, our (fixed) spatial/mass resolution only allows us to treat galaxy formation processes (i.e. star formation, feedback from active galactic nuclei) with the use of sub-grid modelling. While the impact of galaxy formation onto the large-scale dynamics of cosmic plasma have been calibrated and tested in previous works \citep[e.g.][]{va17cqg,gv19}, the role of galaxy evolution in shaping our observables (most noticeably the formation of neutral Hydrogen in this case) shall be further explored in the future with higher resolution simulations. However, the fact that our analysis considered also masked maps (Sec.3.1) allows us to bracket the uncertainties related the densest part of the distribution of cosmic baryons, and by-pass the intrinsic limitations of missing galaxy formation physics.

Our adopted fixed resolution also limits the 
development of a small-scale magnetic dynamo, which is instead predicted by a few works \citep{ry08}. 
While previous dedicated resolution tests 
have found no indications for an increased magnetisation of filaments going to even higher resolution, and have proposed physical reasons for the lack of dynamo amplification in cosmic filaments \citep[][]{va14mhd,gh16}, our  sub-grid dynamo model (DYN5) is explicitly designed to consider the impact of unresolved dynamo amplification in our final magnetic field model. Limited to the comparison with the recent results of \citet{vern17}, the DYN5 model seems to produce a too large average magnetisation of filaments in the cosmic web to be a viable model.

Third, our synthetic observations contain gross oversimplifications concerning the generation of noise in real observations, which has subtleties and different features at different wavelengths. For example, the role of the Galactic Foreground in synchrotron emission and Faraday Rotation is here neglected, as well as the contamination from (pointlike or extended) radio galaxies. On the other hand, the role of the particle and instrumental background on X-ray emission is also treated in an approximate way, by incorporating their effects into our definition of detection threshold. 

Finally, we did not produce proper lightcones for the various observables (e.g. by stacking several simulated boxes along the line of sight up to a large redshift), but, for the sake of simplicity, we limited our first analysis to a local volume at low redshift.

Given the above limitations, our  results suggest that the statistical correlation between the galaxy network and radio observables is a promising tool to probe the amplitude of extragalactic magnetic fields,  well outside of the cluster volume usually explored by existing radio observations, allowing to discriminate between competing models of magnetogenesis. 

Observable proxies related to the thermal properties of the gas are also well correlated with the galaxy distribution out to several $\rm Mpc$, with relatively little dependence on the assumed baryonic physics.
We conclude that with presently available surveys of galaxies and with current multi-wavelength instruments, the best chances to detect the diffuse IGM outside of halos via cross-correlation are given by SZ observations. 

An additional interesting probe is represented by the correlation between the galaxy distribution and surveys of Rotation Measure and synchrotron emission. The cross-correlated signal appears to be detectable already with current facilities if the magnetic field in filaments is volume filling and of the order of $\geq 1-10 \rm ~nG$, as expected in our primordial or in the dynamo scenario.
However, our first test to mimic the cross-correlation between the WISE/2MASS survey and MWA Phase I observation as in \citet{vern17} suggests that $\leq 10 ~\rm nG$ magnetic fields are required in filaments not to violate observational constraints.
This can be accomplished in a primordial magnetogenesis scenario with a primordial magnetic fields os strength $\leq 0.1 \rm ~nG $, or by an astrophysical origin of extragalactic magnetic fields, which would produce little volume filling magnetic fields in the cosmic web. 

With future work, we plan to improve on some of the limitations outlined above, i.e. by modelling longer lightcones and including more realistic templates of noise in synthetic observations, which can be crucial in order to quantitatively assess the robustness of possible (albeit marginally significant) detections of diffuse emission from the cosmic web in existing and future cross-correlation studies. In the meantime, the main results of our current cross-correlation analysis, for different models and observables, are publicly available at  https://cosmosimfrazza.myfreesites.net/library-of-cosmic-web-properties.

\section*{Acknowledgements}

The cosmological simulations were performed with the {\enzo} code (http://enzo-project.org), which is the product of a collaborative effort of scientists at many universities and national laboratories. We gratefully acknowledge the {\enzo} development group for providing extremely helpful and well-maintained on-line documentation and tutorials.
F.V. acknowledges financial support from the ERC  Starting Grant "MAGCOW", no. 714196.   
The simulations on which this work is based have been produced on Piz Daint supercomputer at CSCS-ETHZ (Lugano, Switzerland) under projects s701 and s805 and on the J\"ulich Supercomputing Centre (JFZ) under project HHH42 and {\it stressicm}, in numerical projects with F.V. as PI. 
We also acknowledge the usage of online storage tools kindly provided by the INAF Astronomical Archive (IA2) initiative (http://www.ia2.inaf.it).  We thank Marcus Br\"{u}ggen for his support in the first production of the Chronos++ suite of simulations employed in this work,  N. Locatelli and C. Giocoli for useful feedback on future observations.

\bibliographystyle{mnras}
\bibliography{franco,franco2}

\appendix

\end{document}